\documentclass[12pt]{article}

\usepackage{jheppub}
\makeatletter
\def\@fpheader{\relax}
\makeatother

\usepackage{amsthm,amsmath,amssymb,mathtools}
\usepackage{mathrsfs}
\usepackage{bm}
\usepackage{bbm}
\usepackage{graphicx}
\usepackage{float}
\usepackage{tabularx}
\usepackage{caption}
\usepackage{subcaption}
\usepackage[most]{tcolorbox}
\usepackage[mathscr]{euscript}
\usepackage{dsfont}
\usepackage[mathscr]{euscript}
\usepackage{mathrsfs}
\usepackage{hyperref}
\usepackage{tikz}

\newcommand\blfootnote[1]{%
	\begingroup 
	\renewcommand\thefootnote{}\footnote{#1}%
	\addtocounter{footnote}{-1}%
	\endgroup 
}
\newenvironment{nohyphens}{%
	\hyphenpenalty=10000
	\exhyphenpenalty=10000
	\sloppy %
}{\par}

\title{Mellin space reflections, modularity of elliptic Gamma functions and beyond}

\author[a]{Yang Lei\blfootnote{*The authors are ordered purely alphabetically and should all be viewed as the co-first authors. }}
\author[a]{\!\!, Taotao Li}

\affiliation[\,a]{Institute for Quantum Science \& School of Physical Science and Technology,  Soochow University, No.1 Shizi Street, Suzhou 215006, P.R.~China}

\emailAdd{leiyang@suda.edu.cn}
\emailAdd{ttao314159@163.com}

\abstract{\begin{nohyphens}
We study modular transformation formulas from the viewpoint of Mellin space.  
The basic observation is that the functional relation between the Hurwitz zeta function and the polylogarithm can be used to organize modular transformations as reflection identities of Mellin kernels, while the accompanying polynomial anomalies arise from contour deformations.  
We first illustrate this mechanism for the $q$-$\theta$ function and then extend it to the elliptic Gamma function.  
In the latter case, independently Mellin transforming the two elliptic directions leads to a trilinear reflection identity relating the three elliptic Gamma functions appearing in the SL$(3,\mathbb{Z})$ modular formula, while the associated contour deformation reproduces the cubic Bernoulli polynomial. 
Further reflection formulas lead to a weighted $q$-Pochhammer type function with a modular transformation analogous to that of the $q$-$\theta$ function, as well as a transposed trilinear reflection identity analogous to the Mellin space structure underlying the SL$(3,\mathbb{Z})$ transformation.  
Our results suggest that Mellin space reflection identities provide a useful organizing principle for constructing and studying modular special functions beyond the standard multiple elliptic Gamma hierarchy.

	\end{nohyphens}
}

\date{}

\begin{document} 
	
	\maketitle
	
\section{Introduction}\label{sec:intro}

The modular properties of two dimensional conformal field theories play a central role in determining the asymptotic density of states at high energy \cite{Cardy:1986ie}. 
A CFT defined on a torus is characterized by its complex structure modulus $\tau$, and its partition function is invariant under modular transformations. 
In particular, under the modular $S$ transformation, one has
\begin{equation}
\mathcal{Z}(\tau,\bar{\tau}) = \mathcal{Z}\left(-\frac{1}{\tau},-\frac{1}{\bar{\tau}}\right).
\end{equation}
For notational simplicity, we will often suppress the dependence on $\bar{\tau}$ in what follows.
This relation connects the low temperature $(\tau \to i\infty)$ and high temperature regimes $(\tau \to 0)$, allowing the high energy density of states and the entropy of the dual black hole saddle to be extracted from the low temperature spectrum \cite{Strominger:1996sh}.

A particularly interesting generalization arises in four dimensional superconformal field theories defined on $S^3\times S^1$, whose supersymmetric partition functions are built from elliptic Gamma functions $\Gamma(z;\tau,\sigma)$.
The elliptic Gamma function is the contribution of a four dimensional $\mathcal{N}=1$ chiral multiplet and serves as the fundamental building block of the superconformal index \cite{Romelsberger:2005eg,Dolan:2008qi}. 
The $\Gamma(z;\tau,\sigma)$ satisfies the SL$(3,\mathbb{Z})$ modular relation \cite{Felder_2000}:
\begin{align}\label{eq:def-Sl3Z-ellipticGamma}
	\begin{split}
\Gamma(z;\tau,\sigma) &=  e^{-i\pi Q(z;\tau,\sigma)}\Gamma\left(\frac{z}{\sigma};\frac{\tau}{\sigma},-\frac{1}{\sigma}\right)\Gamma\left(\frac{z}{\tau};\frac{\sigma}{\tau},-\frac{1}{\tau}\right) \\
	Q(z;\tau,\sigma) &= \frac{z^3}{3\tau\sigma} -\frac{\tau+\sigma-1}{2\tau\sigma} z^2 + \frac{\tau^2+\sigma^2 +3\tau \sigma -3\tau-3\sigma+1}{6\tau\sigma} z \\
&+ \frac{(\tau+\sigma-1)(\tau^{-1}+\sigma^{-1}-1)}{12} \,.
	\end{split}
\end{align}
Equation \eqref{eq:def-Sl3Z-ellipticGamma} may be viewed as a higher rank analogue of the two dimensional modular relation, since both moduli $\tau$ and $\sigma$ are transformed nontrivially.
This modular property has played a crucial role in extracting the asymptotic growth of supersymmetric states and in understanding the microscopic origin of the entropy of supersymmetric AdS$_5$ black holes \cite{Cabo-Bizet:2018ehj,Choi:2018hmj,Benini:2018ywd,Goldstein:2020yvj,Gadde:2020bov}. 
Relations of the form \eqref{eq:def-Sl3Z-ellipticGamma} are closely related to holomorphic factorization \cite{Nieri:2015yia,Jejjala:2022lrm}, whose geometric origin has also been related to the Heegaard splitting of three dimensional manifolds \cite{Gadde:2020bov,Jejjala:2022lrm}.

Unlike the two dimensional case, the modularity associated with the elliptic Gamma function is closely related to the supersymmetry. 
It is therefore natural to ask whether analogous modular structures also exist in non-supersymmetric higher dimensional quantum field theories.
One class of examples consists of conformal field theories defined on $T^d$, whose partition functions exhibit an SL$(d,\mathbb{Z})$ modular symmetry inherited from the mapping class group of the torus \cite{Shaghoulian:2015kta,Alessio:2021krn}.
In this case, the modular transformations simply permute the $S^1$ cycles and therefore admit a clear geometric interpretation.

A more nontrivial class arises for conformal field theories on $S^{d-1}\times S^1$ studied earlier in \cite{Cappelli:1988vw}, where no obvious geometric modular group exists.
This direction was pioneered by Cardy \cite{cardy1991operator} and subsequently developed in \cite{Kutasov:2000td,Lei:2024oij} by constructing reflection invariant quantities in the Mellin space of the partition function \footnote{See also \cite{Shaghoulian:2016gol} for an intriguing modular relation involving a change of the background topology.}.
However, in these examples, the partition function itself is generally not the object acted by modular transformation. 
Instead, Cardy showed that for the four dimensional free scalar theory, the Casimir energy function $E_4(\tau) = \frac{d}{d\tau} \ln \mathcal{Z}_4(\tau)$ 
transforms as a modular form of weight four \cite{cardy1991operator}. 
This obscures the physical interpretation of modularity, since the modular transformation no longer acts directly on the partition function itself.

A different perspective was proposed in \cite{Lei:2024oij}. 
Rather than seeking a geometric origin of modularity, one may start from a partition function determined by the physical spectrum and ask what is the most natural modular structure associated with the corresponding special function. 
From this viewpoint, it was shown that the partition functions of free scalar theories on even dimensional manifolds are naturally identified with higher rank elliptic Gamma functions, transforming under the SL$(d-1,\mathbb Z)$ modular group \cite{NARUKAWA2004247}.
Such modular structures are potentially useful not only for extracting the leading Cardy-like growth, but also for organizing the subleading terms in the high temperature expansion.
For free conformal theories, refined asymptotic analyses based on Hilbert series and generalized Meinardus' theorem determine both the leading exponential growth and nontrivial subleading corrections and multiplicative coefficients \cite{Melia:2020pzd}.
More generally, thermal effective field theory organizes the high temperature expansion of CFT partition functions as a derivative expansion, whose subleading coefficients encode additional CFT data \cite{Benjamin:2023qsc,Allameh:2024qqp,Benjamin:2024kdg}.

While the Mellin transformation approach provides an alternative framework for uncovering hidden modular structures, several challenges remain.
The existing construction is most effective for unrefined partition functions depending on a single chemical potential. 
In contrast, supersymmetric partition functions in higher dimensions generally depend on multiple angular momenta and $R$-symmetry fugacities.
Extending the Mellin approach to such refined partition functions naturally leads to multivariable (inverse) Mellin transforms, whose analytic structure is considerably more involved. 
See \cite{Penedones:2019tng} for related applications of multivariable
Mellin representations to conformal correlation functions.
A second limitation is that the reflection invariant quantity constructed in Mellin space does not always correspond to a local operation on the original partition function. 
In certain examples, such as the three dimensional free scalar theory \cite{cardy1991operator, Oshima:1992yy}, the modular object is obtained only after applying a non-local transformation to the partition function, obscuring its physical interpretation.

The use of Mellin-(Barnes) transformations in the study of special functions, functional equations, and asymptotic expansions has a long history, see \cite{Paris_Kaminski_2001} for a systematic review.
Particularly relevant to the present work is the framework systematically
developed and reviewed by \cite{doi:10.1142/8711},
where functional equations of zeta functions are regarded as a guiding principle for constructing a broad class of modular relations through Mellin transforms, special function kernels, and contour deformations.
Importantly, the framework does not require the reflected Dirichlet series to
coincide with the original one. 
A typical Riemann type functional equation considered there takes the form 
\begin{equation}
\Gamma(r-s) \psi(r-s) = \Gamma(s) \phi(s)
\end{equation}
where $\phi$ and $\psi$ may be different Dirichlet series.
Consequently, the function obtained after the modular transformation need not be identical to the original one.
Many of the corresponding modular kernels in \cite{doi:10.1142/8711} are naturally expressed in terms of Fox $H$-functions and Meijer $G$-functions.

The extension from one to several independent Mellin variables, however, is
analytically nontrivial.
In multiple Mellin-Barnes integrals, pole loci associated with one integration
variable may depend on the others, so that contour deformations naturally
encounter linked pole hyperplanes and leave lower dimensional residual integrals \cite{Paris_Kaminski_2001}.
Moreover, convergence itself is genuinely multidimensional: a double Mellin-Barnes integral may fail to converge even when the corresponding iterated one dimensional integrals are individually convergent \cite{doi:10.1142/1425}.

A further motivation of the present work concerns the uniqueness of the SL$(3,\mathbb Z)$ modular structure. 
At present, the elliptic Gamma function \cite{Felder_2000,Nishizawa_2001} appears to be the prototypical example of a special function satisfying the modular relation \eqref{eq:def-Sl3Z-ellipticGamma} together with other identities of SL$(3,\mathbb{Z}) \ltimes \mathbb{Z}^3$. 
This naturally raises the question of whether there exist other families of special functions obeying similar SL$(3,\mathbb Z)$ modular transformation laws, possibly with different phase factors.
For the ordinary SL$(2,\mathbb Z)$ modular group, the construction of modular invariant partition functions is highly systematic. 
Given a collection of conformal characters, one may construct modular invariant combinations by imposing modular invariance,
\begin{equation}
\mathcal Z(\tau,\bar\tau) = \sum_{\mu,\nu} M_{\mu\nu} \, \chi_\mu(\tau)\, \overline{\chi_\nu(\tau)},
\end{equation}
where the matrix $M_{\mu\nu}$ is constrained by the modular $S$ and $T$ transformations \cite{di1997conformal,Cappelli:1986hf}. 
Such a general construction principle, to the best of our knowledge, is presently absent for SL$(3,\mathbb Z)$ modularity.
It is therefore interesting to understand whether the elliptic Gamma function is an isolated example or merely the first member of a broader class of modular special functions. 
Such a framework could provide new modular building blocks for partition functions of quantum field theories and may ultimately admit a geometric interpretation analogous to the two dimensional case.

In this work, we develop a construction of modular reflection
identities in Mellin space.
In Section~\ref{sec:review}, we first revisit the one dimensional prototype, where the Riemann zeta functional equation turns the modular transformation into a reflection of the Mellin integrand followed by a contour deformation. 
In Section~\ref{sec:qthetananalysis}, we extend this mechanism to the $\theta(z;\tau)$ function.  
By organizing the polylogarithm and Hurwitz zeta functions into two component vectors, we show that the modular transformation is encoded by a bilinear scalar reflection identity, conveniently expressed through the componentwise product and counit introduced in Section~\ref{sec:qthetananalysis}.
We further construct a new weighted $q$-Pochhammer type function whose modular $S$-transformation follows from a shifted Hurwitz-polylogarithm reflection identity, closely paralleling the $\theta(z;\tau)$ construction.
In Section~\ref{sec:ellipticGamma}, the same construction is promoted to two Mellin variables and a trilinear reflection.  
The three elliptic Gamma functions in the SL$(3,\mathbb{Z})$ modular formula are unified by a single pointwise identity in Mellin space, while the cubic Bernoulli phase arises from the residues generated when the corresponding Mellin cycles are deformed.  
We also construct a transposed trilinear sector, containing two polylogarithmic
vectors and one Hurwitz zeta vector, which closes algebraically under the same reflection system but has a more subtle realization in the modular variables. 
We then conclude this paper in Section~\ref{sec:conclusion} with discussions and outlooks.

\section{Modular formula and linear Mellin space prototype}
\label{sec:review}

In theories such as $\mathcal{N}=4$ SYM \cite{Aharony:2003sx} and free conformal
field theories, partition functions can often be constructed from the enumeration
of single particle states.
Starting from the single letter partition function
\begin{equation}
\mathcal{Y}(\beta)=\text{tr}\left(e^{-\beta H}\right), \qquad \beta=-2\pi i \tau \,,
\end{equation}
the corresponding multi-letter partition function is obtained by the plethystic
exponential (PE) \cite{Benvenuti:2006qr,Feng:2007ur},
\begin{equation}
\widetilde{\mathcal Z}(\beta) =\text{PE}\!\left[\mathcal{Y}(\beta)\right]=\exp\left[
\sum_{l=1}^{\infty}\frac{1}{l}\mathcal{Y}(l\beta)\right].
\end{equation}
We use $\tilde{Z}(\beta)$ to denote the partition function as the function of inverse temperature $\beta$ and $Z(\tau)$ to denote the partition function as a function of moduli $\tau$. 

In general, $\widetilde{\mathcal Z}(\beta)$ can be a complicated special
function, and its modular properties are not manifest from the plethystic
representation.
The Mellin transform provides a useful way to reorganize the plethystic
exponential.
We will follow \cite{Cardy:1986ie,Gibbons:2006ij} to schematically explain this.
For a theory with spectrum $E_n$ and degeneracy $d_n$, we introduce the Hamiltonian zeta function  $\zeta_H(s) = \sum_n d_n E_n^{-s}$.
It is related to the single-letter partition function by Mellin transformation over $\beta$:
\begin{equation}
\zeta_H(s) = \frac{1}{\Gamma(s)} \int_0^{\infty} \mathcal{Y}(\beta) \beta^{s-1} d\beta \,.
\end{equation}
Using the inverse Mellin transform in the plethystic exponential gives
\begin{equation}\label{eq:Mellin-multiple-def}
\ln \tilde{Z}(\beta)  = \frac{1}{2\pi i} \int_{c-i\infty}^{c+ i \infty}ds\, \beta^{-s} \Gamma(s) \zeta(s+1) \zeta_H(s)  \,,
\end{equation}
where the contour is chosen such that $\text{Re} (s)=c$ lies to the right of all
singularities of the integrand.
The additional factor $\zeta(s+1)$ is the consequence of the plethystic sum over $l$. 
As mentioned in Section \ref{sec:intro}, this representation is particularly useful for deriving modular relations of free conformal field theories in various dimensions \cite{cardy1991operator,Kutasov:2000td}. 
It also provides the natural starting point for applications of Meinardus' theorem \cite{andrews1998theory,Melia:2020pzd,Lucietti:2008cv}, which extracts the asymptotic growth of the corresponding partition functions.

To illustrate the basic mechanism, let us consider the large-$N$ $1$-matrix
partition function as the simplest example \footnote{This terminology follows the  $d$-matrix theory studied in \cite{Lei:2024oij,Collins:2008gc,Bhattacharyya:2008rb,Dolan:2007rq, Lei:2026fep,OConnor:2026zlf,Ramgoolam:2018epz}, with the unrefined partition function of the form $$
\mathcal{Z}_d(q) = \prod_{i=1}^\infty \frac{1}{1-d q^i}  
\,.$$}. 
Writing $q=e^{-\beta} = e^{2\pi i \tau}$, the partition function is
\begin{equation}\label{eq:def-1matrix-parfn}
Z(\tau) = \tilde{Z}(\beta) =	\mathcal{Z}_1(q) = \prod_{i=1}^\infty \frac{1}{1- q^i} \,.
\end{equation}
This function can be also represented as the $q$-Pochhammer symbol $(q;q)_{\infty}^{-1}$. 
It counts multi-trace gauge invariant operators constructed from a single bosonic matrix and is, at the same time, the generating function of integer partitions.
The corresponding single letter spectrum is $E_n=n$ with unit degeneracy, and hence the Hamiltonian zeta function is simply the Riemann zeta function $\zeta(s)$. 

Under the $S$-transformation $\tau\to -\frac{1}{\tau}$, the partition function satisfies
\begin{equation}\label{eq:Stransform-1matrix}
	Z\left( -\frac{1}{\tau}\right) = \frac{1}{\sqrt{\tau}} e^{-i\pi P(\tau)} Z(\tau) \,,\qquad P(\tau) = \frac{1}{12} \left(\tau+\frac{1}{\tau}-3\right) \,.
\end{equation}
where the principal branches of the logarithm and square root are understood for $\operatorname{Im}\tau>0$.
Using \eqref{eq:Mellin-multiple-def} with $\zeta_H(s)=\zeta(s)$, the partition
function admits the Mellin representation
\begin{equation}\label{eq:Mellin-1matrix-parfn}
\ln Z(\tau) = \frac{1}{2\pi i} \int_{c-i\infty}^{c+i\infty} ds \, \frac{1}{(-2\pi i \tau)^s} \Gamma(s) \zeta(s+1) \zeta(s) \,.
\end{equation}
where $c>1$ may be chosen so that the contour lies to the right of all
singularities of the integrand.
For $\operatorname{Im}\tau>0$, its modular image $-1/\tau$ also lies in the
upper half-plane, and therefore
\begin{equation}\label{eq:Mellin-Simage-1matrix-parfn}
\ln	Z\left(-\frac{1}{\tau}\right) = \frac{1}{2\pi i} \int_{c-i\infty}^{c+i\infty}ds \, \frac{\tau^s}{(2\pi i)^s} \Gamma(s) \zeta(s+1) \zeta(s) \,.
\end{equation}
To bring the $S$-transformed expression back to the original Mellin form,
we perform the reflection $s\to-s$, which moves the contour from
$\operatorname{Re}s=c$ to $\operatorname{Re}s=-c$.
Using the Riemann zeta functional equation (see \cite{whittaker1945course, elizalde1994zeta} for comprehensive reviews and generalizations)
\begin{equation}\label{eq:Riemann-reflection}
\zeta(s) = \Gamma(1-s)2^s \pi^{s-1} \sin\left(\frac{\pi s}{2}\right) \zeta(1-s) \,,
	\end{equation}
the integrand of \eqref{eq:Mellin-Simage-1matrix-parfn} can then be shown to be identical to the original Mellin representation \eqref{eq:Mellin-1matrix-parfn},
\begin{equation}
\left(\frac{2\pi i}{\tau}\right)^s \Gamma(-s)\zeta(1-s)\zeta(-s) = \frac{1}{(-2\pi i\tau)^s} \Gamma(s)\zeta(s+1)\zeta(s).
\end{equation}
Thus the Mellin integrand is invariant under the combined modular and Mellin
reflections; only the integration contour has changed.

It follows that
\begin{equation}\label{eq:oint-1matrix}
\ln Z(\tau) - \ln Z \left(-\frac{1}{\tau}\right) = \frac{1}{2\pi i } \oint_{\mathcal{C}} ds \frac{1}{(-2\pi i \tau)^s} \Gamma(s) \zeta(s+1) \zeta(s) \,,
\end{equation}
where $\mathcal C$ encloses the strip between $\operatorname{Re}s=-c$ and $\operatorname{Re}s=c$.
The apparent poles of $\Gamma(s)$ at negative integers are almost entirely
cancelled by the trivial zeros of the zeta functions.
At $s=-2,-4,\ldots$, the poles are cancelled by the zeros of $\zeta(s)$,
while at $s=-3,-5,\ldots$, they are cancelled by the zeros of $\zeta(s+1)$.
The only remaining singularities are therefore at $s=-1,0,1$, with $s=0$ being a double pole. 
Their residues then add up to be 
\begin{align}
\left(	\text{Res}_{s=-1} + \text{Res}_{s=0} + 	\text{Res}_{s=1} \right) \left[\frac{1}{(-2\pi i \tau)^s} \Gamma(s) \zeta(s+1) \zeta(s) \right] = i\pi P(\tau) + \frac{1}{2}\ln \tau \,.
\end{align}
Equation \eqref{eq:oint-1matrix} therefore reproduces precisely the modular
relation \eqref{eq:Stransform-1matrix}.

More generally, the framework reviewed in \cite{doi:10.1142/8711} relates functional equations in Mellin space to modular relations after Mellin inversion and contour deformation.
Although Mellin representations such as \eqref{eq:Mellin-multiple-def} are often straightforward to obtain, identifying the appropriate reflection structure that closes the transformed Mellin kernel is considerably less systematic.
In the following sections, we generalize this mechanism first to the
$q$-$\theta$ function and then to the elliptic Gamma function, where the linear
reflection considered above is replaced by bilinear and trilinear reflection
structures, respectively.

\section{$q$-$\theta$ function from the bilinear Mellin reflection}
\label{sec:qthetananalysis}

\subsection{Mellin space and modularity}
\label{ssec:modularity-qtheta}

The $q$-$\theta$ function $\theta(z;\tau)$, with
$x=e^{2\pi i z}$ and $q=e^{2\pi i\tau}$, appears as the elliptic genus of a
two dimensional $\mathcal N=(0,2)$ chiral multiplet
\cite{Benini:2013nda,Kawai:1994np}, as well as the partition function of a
four dimensional $\mathcal N=1$ chiral multiplet on $T^2\times S^2$
\cite{Closset:2013sxa,Gadde:2020bov}.
It admits both the product and plethystic representation,
\begin{align}
	\begin{split}
\theta(z;\tau) = \prod_{n=0}^\infty (1-x q^n) \left( 1-\frac{q^{n+1}}{x}\right) = \exp\left[- \sum_{l=1}^\infty \frac{1}{l} \frac{x^l + q^{l}x^{-l}}{1-q^l}\right] \,.
	\end{split}
\end{align}
The function is invariant under the $T$ transformation
$\tau\to\tau+1$, while under the $S$ transformation it satisfies
\begin{align}\label{eq:Stransform-qtheta}
	\begin{split}
& \theta(z;\tau) = e^{-i\pi B(z;\tau)} \theta \left( \frac{z}{\tau};-\frac{1}{\tau}\right) \\
& B(z;\tau) = \frac{z^2}{\tau} +z\left( \frac{1}{\tau}-1\right) + \frac{1}{6} \left(\tau+\frac{1}{\tau}\right) -\frac{1}{2} \,.
	\end{split}
\end{align}
Together with integer shifts of $z$, the $S$ and $T$ transformations generate
the SL$(2,\mathbb Z)\ltimes\mathbb Z^2$ modular structure of
$\theta(z;\tau)$.
Modular formulae associated with more general elements of
SL$(2,\mathbb Z)$ can be found in \cite{Felder_2008,Jejjala:2022lrm}.

Because of the additional parameter $z$, the Mellin representation of
$\theta(z;\tau)$ is not unique.
To represent the modular relation \eqref{eq:Stransform-qtheta} as a reflection
formula in Mellin space, it is convenient to parametrize $z=u+v\tau$, namely, to expand $z$ in a basis adapted to the torus lattice.
For definiteness, we restrict to $0<u,v<1$ so that the parameters stay away from the singular loci relevant below.
Using the unit periodicity of the $q$-$\theta$ function, we may equivalently write
the transformed coordinate as 
\begin{equation}
	\frac{z-1}{\tau} = v + (1-u) \left(-\frac{1}{\tau}\right) \,.
\end{equation}
The modular relation can therefore be expressed as
\begin{equation}\label{eq:modular-S-QTHETA}
\theta(z;\tau) = e^{-i\pi B(z-1;\tau)} \theta \left(\frac{z-1}{\tau};-\frac{1}{\tau}\right)  \,,
\end{equation}
and the $S$-transformation induces the simple map
\begin{equation}
(u,v;\tau) \to \left(v,1-u;-\frac{1}{\tau} \right) \,.
\end{equation} 

We next construct a Mellin representation adapted to the coordinates
$(u,v)$.
In contrast to the 1-matrix example of Section~\ref{sec:review}, the plethystic sum no longer produces a single Riemann zeta factor $\zeta(s+1)$.
Consider first the $q$-Pochhammer symbol $(x;q)_\infty$. Its logarithm can be
written as
\begin{equation}
\ln (x;q)_\infty = -\sum_{l=1}^\infty \frac{e^{2\pi i l u}}{l} \sum_{n=0}^\infty e^{2\pi i(n+v)l\tau} \,.
\end{equation} 
Applying the Cahen-Mellin representation
\begin{equation}\label{eq:Cahen--Mellin}
e^{-x} = \frac{1}{2\pi i} \int_{c-i\infty}^{c+i\infty} ds \, \Gamma(s) x^{-s} \,,
\end{equation} 
the double sum can be reorganized as
\begin{equation}
\ln (x;q)_\infty = -\frac{1}{2\pi i} \int_{c-i\infty}^{c+i \infty}  ds \, \frac{1}{(-2\pi i \tau)^s} \Gamma(s) \text{Li}_{1+s} (e^{2\pi i u}) \zeta(s,v) \,,
\end{equation}
Here $\operatorname{Li}_s$ and $\zeta(s,v)$ denote the polylogarithm and Hurwitz zeta function respectively,
\begin{equation}
\text{Li}_s (x) = \sum_{n=1}^\infty \frac{x^n}{n^s}, \qquad \zeta(s,v) = \sum_{n=0}^\infty \frac{1}{(n+v)^s} \,,
\end{equation}
with analytic continuation understood outside their domains of absolute
convergence.
Combining the two $q$-Pochhammer factors, the Mellin representation of the
$q$-$\theta$ function becomes
\begin{align}\label{eq:Mellin-qtheta}
	\begin{split}
\ln \theta(z;\tau) &= - \frac{1}{2\pi i} \int_{c-i\infty}^{c+i \infty} ds \, \frac{1}{(-2\pi i \tau)^s}  \Gamma(s)  \\
& \times \left[\text{Li}_{s+1}(e^{2\pi i u}) \zeta(s,v) + \text{Li}_{s+1}(e^{2\pi i (1-u)}) \zeta(s,1-v) \right] \,.
\end{split}
\end{align}

To make the reflection structure of \eqref{eq:Mellin-qtheta} manifest, it is useful to introduce two-component vectors \footnote{The choice of minus sign in the second component is for adapting to the multiple elliptic Gamma function, as we will see in Section \ref{sec:ellipticGamma} and appendix \ref{appendix}. }
\begin{equation}\label{eq:def-mathbfLH}
\mathbf{L}_s(u) = \left(
\begin{array}{c}
	\text{Li}_{s+1} (e^{2\pi i u})\\
	-\text{Li}_{s+1} (e^{2\pi i (1-u)})
\end{array}
\right)\,, \qquad 
\mathbf{H}_s(v) = \left(
\begin{array}{c}
	\zeta(s,v)\\
	-\zeta(s,1-v)
\end{array}
\right)  \,.
\end{equation}
which we regard as elements of a two dimensional vector space $V$.
For later generalization, we equip $V$ with the componentwise multiplication
and the linear functional
\begin{align}
	\begin{split}
\text{multiplication}\qquad &\circ:\, V\times V \to V, \\
\text{counit} \qquad &   \varepsilon:\, V\to \mathbb{C}
	\end{split}
\end{align}
defined by 
\begin{equation}
\begin{pmatrix}a_1\\a_2\end{pmatrix}\circ
\begin{pmatrix}b_1\\b_2\end{pmatrix} =
\begin{pmatrix}a_1b_1\\a_2b_2\end{pmatrix}, \qquad
\varepsilon
\begin{pmatrix}a_1\\a_2\end{pmatrix} = a_1+a_2.
\end{equation}
These operations provide $V$ with a simple realization of a commutative
Frobenius algebra, with $\varepsilon$ playing the role of the counit
\cite{kock2004frobenius}.
The associated Frobenius pairing is
\begin{equation}
	\langle \mathbf a,\mathbf b\rangle
	\equiv
	\varepsilon(\mathbf a\circ\mathbf b).
\end{equation}
We therefore define the following scalar Mellin kernel:
\begin{equation}\label{eq:qtheta-bilinear-kernel}
\Xi_2(u,v;s) \equiv \frac{\Gamma(s)}{(2\pi)^s} \, \varepsilon\left[ \mathbf L_s(u)\circ\mathbf H_s(v) \right].
\end{equation}
such that the $\theta(z;\tau)$ can be compactly organized as 
\begin{equation}\label{eq:Mellin-qtheta-Xi}
\ln\theta(z;\tau) = -\frac{1}{2\pi i}
\int_{c-i\infty}^{c+i\infty} ds\, (-i\tau)^{-s}\, \Xi_2(u,v;s).
\end{equation}
Similarly, using the $S$-transformed coordinates
$(u,v)\to(v,1-u)$, we obtain
\begin{equation}\label{eq:Mellin-qtheta-S-Xi}
\ln\theta\left( \frac{z-1}{\tau};-\frac{1}{\tau} \right) = -\frac{1}{2\pi i} \int_{c-i\infty}^{c+i\infty} ds\, \left(\frac{i}{\tau}\right)^{-s} \Xi_2(v,1-u;s).
\end{equation}

As in $1$-matrix theory, in order to compare the Mellin integrands before and after the modular transformation, we perform the reflection $s\to -s$.
The essential ingredient is the functional relation between the Hurwitz zeta
function and the polylogarithm \cite{komori2010barnes,whittaker1945course},
\begin{align}\label{eq:Hurwitz-Polylog-reflection}
\zeta(s,v) = \frac{\Gamma(1-s)}{(2\pi)^{1-s}} \left( e^{\frac{\pi i(s-1)}{2}} \text{Li}_{1-s}(e^{2\pi iv}) + e^{\frac{\pi i(1-s)}{2}} \text{Li}_{1-s} (e^{2\pi i(1-v)}) 
\right)  \,.
\end{align}
For $0<v<1$, this relation follows directly from the Fourier expansion for $\operatorname{Re}(s)<0$ and is understood elsewhere by analytic continuation.
In terms of the vector notation introduced above, the corresponding reflection
relations take the compact form
\begin{align}\label{eq:reflectionLH-eachlinear}
	\begin{split}
\mathbf{L}_{-s} (u) & = \frac{\Gamma(s)}{(2\pi)^s} \mathbf{M}(s) \cdot \mathbf{H}_s(u)\,, \qquad 
\mathbf{H}_{-s} (1-v)  = \frac{\Gamma(1+s)}{(2\pi)^{1+s}} \mathbf{M}(1+s) \cdot \mathbf{L}_s(v) \,,
	\end{split}
\end{align} 
where connection matrix $\mathbf M(s)$ acts by ordinary matrix multiplication and is given by
\begin{equation}
	\mathbf{M}(s) = \left(
	\begin{array}{cc}
e^{\frac{\pi i s}{2}}		& -e^{-\frac{\pi i s}{2}}  \\
-e^{-\frac{\pi i s}{2}}		&  e^{\frac{\pi i s}{2}}
	\end{array}
	\right) \,.
\end{equation} 
An important feature of \eqref{eq:reflectionLH-eachlinear} is that neither
$\mathbf{L}$ nor $\mathbf{H}$ closes under the reflection separately.
Instead, the reflection exchanges the two types of functions.
Nevertheless, their Frobenius pairing introduced in \eqref{eq:qtheta-bilinear-kernel} closes into itself, satisfying
\begin{equation}\label{eq:reflection-Xi2}
	\Xi_2(u,v;s)=\Xi_2(v,1-u;-s)\,.
\end{equation}
by using \eqref{eq:reflectionLH-eachlinear}.
Thus, although the individual linear components are exchanged by reflection,
their bilinear combination defines a scalar reflection invariant.

We are now ready to establish the connection between the Mellin reflection
formula and the modular transformation of the $q$-$\theta$ function.
After performing the reflection $s\to -s$ and using \eqref{eq:reflection-Xi2}, the Mellin integrands in \eqref{eq:Mellin-qtheta-Xi} and \eqref{eq:Mellin-qtheta-S-Xi} become identical. 
The only remaining difference is the position of the integration contour. 
Their difference is therefore given by a contour integral enclosing the poles crossed during the deformation,
\begin{equation}\label{eq:qtheta-contour}
\ln \theta(z;\tau) - \ln \theta\left(\frac{z-1}{\tau}; -\frac{1}{\tau} \right) = - \frac{1}{2\pi i} \oint_{\mathcal{C}} ds \, (-i\tau)^{-s}\Xi_2(u,v;s)\,.
\end{equation}
Due to $u \in(0,1)$, the polylogarithm function Li$_{s+1}(e^{2\pi iu})$ is regular in $s$. 
The possible singularities of the integrand arise from the poles of $\Gamma(s)$ at $s=-m$, $m\geq0$, together with the pole of the Hurwitz zeta function at $s=1$.
Most of the apparent poles at negative integers, however, are cancelled.
At non-negative integers $m$, the Hurwitz zeta function satisfies \cite{elizalde1994zeta}
\begin{equation} \label{eq:Hurwitzzeta-to-Bernoulli}
	\zeta(-m,v) = - \frac{B_{m+1}(v)}{m+1}\,,
\end{equation}
where the ordinary Bernoulli polynomial $B_m(v)$ is subject to $B_{m}(1-v)= (-1)^{m} B_m(v)$. 
It follows that $\zeta(-m,1-v)= (-1)^{m+1} \zeta(-m,v) $. 
On the other hand, the polylogarithms at these integer values satisfy
\begin{equation}\label{eq:Linega-identity-m}
	\text{Li}_{1-m} (e^{2\pi iu}) + (-1)^{m+1}	\text{Li}_{1-m} (e^{2\pi i(1-u)}) =0, \qquad m\ge 2\,.
\end{equation}
Consequently, the zeros of the bilinear combination cancel the poles of
$\Gamma(s)$ for all $s=-m$ with $m\geq2$.
The only remaining singularities are therefore $s=0,\pm 1$.
The corresponding residues are
\begin{align}
	\begin{split}
& \text{Res}_{s=1} \left[ -(-i\tau)^{-s}\Xi(u,v;s)\right] =- \frac{i\pi}{\tau} \left(u^2-u+\frac{1}{6}\right) \\
& \text{Res}_{s=0} \left[ -(-i\tau)^{-s}\Xi(u,v;s)\right] = \pi i\left(v-\frac{1}{2}\right) (1-2u) \\
& \text{Res}_{s=-1} \left[ -(-i\tau)^{-s}\Xi(u,v;s)\right] = - i\pi \left(v^2-v+\frac{1}{6}\right) \tau \,.
	\end{split}
\end{align}
Their sum reproduces precisely the polynomial phase appearing in the $q$-$\theta$ modular transformation $-i\pi B(z-1;\tau)$, as \eqref{eq:Stransform-qtheta}. 

Equation \eqref{eq:qtheta-contour} therefore gives \eqref{eq:modular-S-QTHETA}, showing that the bilinear reflection identity \eqref{eq:reflection-Xi2} is the Mellin space counterpart of the modular $S$-transformation of the $\theta(z;\tau)$ function.
More general SL$(2,\mathbb Z)$ transformations can be generated using multiplication formulae \cite{Felder_2000,felder2002multiplication,Jejjala:2022lrm}, which we will not pursue here.

The above derivation also suggests a useful way of organizing modular transformations in Mellin space. 
The modular relation naturally separates into two ingredients: a reflection symmetry of the Mellin kernel and the residues generated by the associated contour deformation. 
The former captures the invariant structure, while the latter produces the polynomial phase accompanying the modular transformation. 
This separation provides a useful strategy for identifying modular structures of more general special functions.
A natural question is then whether the reflection function $\Xi(u,v;s)$ is uniquely determined by the underlying linear reflection relations. 
We address this question in the following subsection.

\subsection{Bilinear reflection invariant in linear space}
\label{ssec:algebratoqtheta}

In the previous subsection, we showed that a particular bilinear combination of the Hurwitz zeta and polylogarithm functions plays a central role in the Mellin space description of the modular transformation of the $q$-$\theta$ function. 
A natural question is how strongly this particular bilinear contraction is
constrained by the underlying linear reflection relations.

There are two conceptually different ways to generalize the construction.
One may either change the assignment of Mellin parameters carried by
$\mathbf{L}_{f(s)}$ and $\mathbf{H}_{g(s)}$, or keep the equal Mellin assignment and vary the bilinear contraction itself.  
In this subsection we consider only the second possibility.  
Namely, we restrict to the equal Mellin variable sector $\mathbf{L}_s \otimes \mathbf{H}_s$, while more general Mellin parameter assignments will be discussed in the next subsection.

The Frobenius pairing used above can equivalently be written as
$\varepsilon(\mathbf a\circ\mathbf b) =\mathbf a^{T}\mathbf b$.
We therefore replace it by a general $s$-independent bilinear form $C\in{\rm End}(V)$ and define $\mathcal K_C(u,v;s) \equiv\mathbf{L}_s^{T}(u)\, C\, \mathbf H_s(v)$.
We ask for which constant matrices $C$ this scalar kernel closes under the same basic reflection
\begin{equation}\label{eq:KC-reflection-condition}
\mathcal K_C(u,v;s)\,\rho(s)= \mathcal K_C(v,1-u;-s),
\end{equation}
where the scalar factor $\rho(s)$ is allowed to depend on the Mellin variable.
More general reflection relations involving additional shifts or transformations of $(u,v)$ may reasonably be considered as alternative possibilities.
Such shifts correspond to lattice translations of $z$ by integers or integer multiples of $\tau$ and generally modify the $q$-$\theta$ function only by known quasi-periodic factors. 
Since our purpose here is to identify the bilinear structure selected by the basic $S$ reflection, we restrict ourselves to the transformation $(u,v)\mapsto(v,1-u)$.

Using the linear reflection relation \eqref{eq:reflectionLH-eachlinear}, together with the symmetry $\mathbf{M}^T(s)=\mathbf{M}(s)$, the closure condition \eqref{eq:KC-reflection-condition} reduces to the purely algebraic intertwining equation
\begin{equation}\label{eq:C-intertwiner}
\mathbf M(s+1)\,C^T\,\mathbf M(s) = \lambda(s)\,C ,
\end{equation}
where $\lambda(s)$ absorbs the scalar factors associated with the two linear reflections.
Since $V$ is two dimensional, a general constant matrix can be expanded in the Pauli basis, $C=\sum_{\mu=0}^{3}\alpha_\mu\sigma_\mu$,  where $\sigma_0$ denotes the identity matrix. 
This parametrization does not impose any restriction on the bilinear form and is particularly convenient to adapt to the reflection matrix which takes the simple form
\begin{equation}\label{eq:form-MMatrix}
\mathbf M(s)= e^{\frac{\pi i s}{2}}\sigma_0- e^{-\frac{\pi i s}{2}}\sigma_1 \,.
\end{equation}
Substituting the ansatz into \eqref{eq:C-intertwiner}  gives
\begin{align}\label{eq:C-Pauli-equations}
	\begin{split}
\lambda(s) &= -2\sin(\pi s),\\
\lambda(s)\alpha_2 &=-2i\cos(\pi s)\alpha_2+2\alpha_3,\\
\lambda(s)\alpha_3&= 2i\cos(\pi s)\alpha_3+2\alpha_2.
	\end{split}
\end{align}
Here the first equation follows from either of the $\sigma_0$ and $\sigma_1$ sectors whenever the corresponding coefficient is nonzero. 
Remarkably, both sectors have the same eigenvalue $\lambda(s)=-2\sin(\pi s)$.
The non-vanishing pair $(\alpha_2,\alpha_3)$ are subject to $\alpha_3=i e^{i\pi s} \alpha_2$, which is incompatible with the assumption that $C$ is independent of $s$. 
Hence the complete space of constant bilinear forms closing under the reflection is
\begin{equation}\label{eq:C-general-solution}
C=\alpha_0\sigma_0+\alpha_1\sigma_1,\qquad\lambda(s)=-2\sin(\pi s).
\end{equation}
Thus, the reflection algebra reduces the four dimensional space of general
bilinear forms on $V$ to the two dimensional subspace spanned by
$\sigma_0$ and $\sigma_1$.

For either allowed bilinear form, the reflected kernel satisfies
\begin{equation}
	\mathcal{K}_C(v,1-u;-s)
	=
	\lambda(s)\,\frac{\Gamma(s) \Gamma(s+1)}{(2\pi)^{2s+1}}\
	\mathcal{K}_C(u,v;s),
\end{equation}
where $\Gamma(s)$ originates from the reflection formula of \eqref{eq:Hurwitz-Polylog-reflection}.
Using the Gamma-function reflection formula, the scalar prefactor can be simplified as
\begin{align}
-2\sin(\pi s) \frac{\Gamma(s)\Gamma(1+s)}{(2\pi)^{2s+1}}=
\frac{\Gamma(s)}{\Gamma(-s)(2\pi)^{2s}} \,.
\end{align}
This suggests introducing the completed bilinear kernel
\begin{equation}\label{eq:def-completed-XiC}
\Xi_2^{(C)}(u,v;s)\equiv\frac{\Gamma(s)}{(2\pi)^s}\mathcal K_C(u,v;s)\,.
\end{equation}
which obeys the reflection formula: $ \Xi^{(C)}_2(u,v;s)= \Xi^{(C)}_2(v,1-u;-s)$.
The Frobenius contraction used for the $q$-$\theta$ function corresponds
to the particular choice $C=\sigma_0$.
The other basis element is related to it by the elementary complement
transformation,
\begin{equation}\label{eq:sigma1-branch}
\Xi_2^{(\sigma_1)}(u,v;s)=-\Xi_2^{(\sigma_0)}(u,1-v;s).
\end{equation}
Thus the two basis elements $\sigma_0$ and $\sigma_1$ are exchanged, up
to a sign, by $v\rightarrow1-v$.  
A generic linear combination $\alpha_0\sigma_0+\alpha_1\sigma_1$, however, remains a genuine element of the two dimensional reflection space and need not be reducible to the pure $\sigma_0$ branch by this discrete transformation.

Within the equal Mellin sector and under the assumption of an $s$-independent bilinear form, the above analysis exhausts the possible bilinear contractions compatible with the reflection algebra. 
The allowed space is the two dimensional subspace spanned by $\sigma_0$ and $\sigma_1$.  
The ordinary $q$-$\theta$ function realizes the Frobenius branch $C=\sigma_0$.  
Once the Mellin parameter assignment is relaxed, further reflection sectors may appear, as we illustrate in the next subsection.

\subsection{An example of shifted Mellin subsector}
\label{ssec:newSL2Zfunction}

The discussion in the previous subsection was restricted to the equal Mellin sector, in which the polylogarithmic and Hurwitz zeta vectors carry the same Mellin label. 
One may instead consider different assignments of the Mellin parameters while keeping the same underlying linear reflection system. 
A particularly simple shifted assignment is
\begin{equation}
\hat{\Xi}_2(u,v;s) = \frac{\Gamma(s)}{(2\pi)^s} \, \varepsilon(\mathbf{L}_{s-1}(u)  \circ\mathbf{H}_{s+1}(v)) \,.
\end{equation}
Compared with the equal Mellin sector $\mathbf{L}_s\circ\mathbf{H}_s$, the Mellin labels of the two factors are shifted by one unit in opposite directions. 
Using the same reflection formula \eqref{eq:Hurwitz-Polylog-reflection} between Hurwitz zeta and polylogarithm, one finds that the completed kernel again closes under reflection,
\begin{equation}\label{eq:Xi-transpose-reflection}
\hat{\Xi}_2 (u,v,s) = \hat{\Xi}_2(v,1-u,-s)\,.
\end{equation}
Thus the reflection algebra admits nontrivial sectors beyond the equal Mellin assignment considered above.

The corresponding function in the original $\tau$ space is defined by the inverse Mellin transform
\begin{equation}
\ln \mathcal{Z} (u,v;\tau) = - \frac{1}{2\pi i} \int_{c-i\infty}^{c+i\infty} ds \, \frac{1}{(- i \tau)^s} \hat{\Xi}_2(u,v,s) \,.
\end{equation}
Let $q=e^{2\pi i\tau}, \,X=e^{2\pi i(u+v \tau)} \,$.
Expanding the polylogarithm and Hurwitz zeta functions in their defining series, the inverse Mellin transform can be written as
\begin{align}\label{eq:Z-transpose-Lambert}
\ln \mathcal{Z}(u,v;\tau) = -\sum_{n=0}^{\infty}
\Bigg[ \frac{1}{n+v} \frac{Xq^n}{1-Xq^n} + \frac{1}{n+1-v} \frac{q^{n+1}/X}{1-q^{n+1}/X} \Bigg].
\end{align}
The new Mellin sector therefore gives rise naturally to a weighted Lambert type series.
To write the function in more compact notation, it is useful to introduce a weighted $q$-Pochhammer symbol by 
\begin{equation} \label{eq:weighted-Pochhammer}
[v,x;q]_\infty = \prod_{n=0}^\infty (1-x q^n)^{\frac{1}{n+v}} \,.
\end{equation}
We then introduce the ratio 
\begin{equation}\label{eq:weighted-theta-ratio}
\Theta(v,X;q) =  \frac{[v,X;q]_\infty}{[1-v,\frac{q}{X};q]_\infty} \,.
\end{equation}
Equation \eqref{eq:Z-transpose-Lambert} can now be expressed compactly as
\begin{equation}
\ln \mathcal{Z}(u,v;\tau) = X\frac{\partial}{\partial X} \ln \Theta(v,X;q) \,.
\end{equation}
Unlike the ordinary $q$-$\theta$ function, its dependence on $u$ and $v$ cannot be reduced to the single combination $z=u+v\tau$.

The reflection identity \eqref{eq:Xi-transpose-reflection} immediately
induces a nontrivial modular $S$ transformation. 
Under $S: (u,v;\tau) \longrightarrow \left(v,1-u;-\frac{1}{\tau}\right)$, 
the reflected Mellin contour can be brought back to the original one.
In contrast to the ordinary $q$-$\theta$ sector, the resulting residue
structure is particularly simple. 
At the negative integers $s=-m$ with $m\geq1$, using \eqref{eq:Linega-identity-m},
\eqref{eq:Hurwitzzeta-to-Bernoulli}, together with the reflection property of the Bernoulli polynomials discussed in Section~\ref{ssec:modularity-qtheta}, the two components of the bilinear kernel cancel. 
Consequently, all poles of $\Gamma(s)$ at negative integers are removed, and the only singularity contributing to the contour deformation is the double pole at
$s=0$.

To see this explicitly, near $s=0$, the functions in the integrand have the expansions
\begin{equation}
\zeta(1+s,a) = \frac{1}{s}-\psi(a)+\mathcal{O}(s),\quad 
\text{Li}_0(e^{\pm 2\pi i u}) = -\frac{1}{2} \pm \frac{i}{2}\cot(\pi u) .
\end{equation}
Since the pole at $s=0$ is of second order, we also need to compute the first order expansion of the polylogarithms.
Differentiating the Hurwitz-polylogarithm reflection formula
at $s=0$ gives
\begin{equation}
\left. \frac{\partial}{\partial s} \left[ \text{Li}_s(e^{2\pi i u})+
	\text{Li}_s(e^{-2\pi i u}) \right] \right|_{s=0} =
-\gamma-\ln(2\pi) -\frac{1}{2} \left[ \psi(u)+\psi(1-u) \right].
\end{equation}
where $\psi(v) = \frac{\Gamma'(v)}{\Gamma(v)}$ is known as the digamma function \cite{whittaker1945course}.
Using the digamma reflection formula $\psi(1-v)-\psi(v)= \pi\cot(\pi v)$, and the remaining Mellin prefactor behaves as
\begin{equation}
\frac{\Gamma(s)}{(2\pi)^s} (-i\tau)^{-s} =
 \frac{1}{s} -\gamma-\ln(-2\pi i\tau) + \mathcal{O}(s)\,.
\end{equation}
We find that the full Mellin integrand has a double pole at $s=0$,
\begin{align}\label{eq:transpose-s0-Laurent}
(-i\tau)^{-s}\hat{\Xi}_2(u,v;s) ={}& -\frac{1}{s^2}
+\frac{\mathcal{A}(u,v;\tau)}{s} +\mathcal{O}(1)\,,
\end{align}
where the residue obtained from these expansions is 
\begin{align}\label{eq:transpose-anomaly}
\mathcal{A}(u,v;\tau) ={}& \ln(-i\tau) +\frac{i\pi}{2}
\cot(\pi u)\cot(\pi v) \nonumber\\
&+ \frac{1}{2} \left[ \psi(v)-\psi(u) +\psi(1-v)-\psi(1-u) \right].
\end{align}
The contour deformation therefore gives 
\begin{align}
\ln \mathcal{Z}\left(v,1-u;-\frac{1}{\tau} \right) - \ln \mathcal{Z}(u,v;\tau) = \mathcal{A} (u,v;\tau) \,.
\end{align}
where $\mathcal{A}$ is the additive modular anomaly in the logarithmic representation.

This example illustrates that changing the Mellin parameter can lead to a genuinely different function in the modular variable $\tau$ while preserving the same reflection pattern. 
The equal sector $\mathbf{L}_s\circ\mathbf{H}_s$ gives the ordinary $q$-$\theta$ function, whereas the shifted sector $\mathbf{L}_{s-1}\circ\mathbf{H}_{s+1}$ leads to the weighted $q$-Pochhammer structure \eqref{eq:weighted-theta-ratio}. 
More general sectors of the form $\mathbf{L}_{f(s)} \circ \mathbf{H}_{g(s)}$ may lead to further bilinear reflection identities and associated special functions. 
We will not pursue such generalizations further in this paper.

\section{Elliptic Gamma function from the trilinear Mellin reflection}
\label{sec:ellipticGamma}

The elliptic Gamma function provides the next natural example beyond the $q$-$\theta$ function considered in the previous section.
It was introduced as an elliptic analogue of the Gamma function and appears in several contexts ranging from integrable lattice models to supersymmetric partition functions \cite{baxter2000partition, Felder_2000, Romelsberger:2005eg, Dolan:2008qi, Gadde:2020bov}.
For the present purpose, its essential feature is that its product and plethystic representations involve two independent elliptic lattice directions, parametrized by the complex moduli $\tau$ and $\sigma$ respectively.
In the supersymmetric field theory realization on a Hopf surface, the corresponding fugacities characterize its complex structure \cite{Assel:2014paa}.
This makes the elliptic Gamma function a natural setting in which the bilinear Mellin reflection of the $q$-$\theta$ function is promoted to a multiple Mellin transform and a trilinear reflection structure.

We begin with the generalized $q$-Pochhammer symbol. 
Introducing $x=e^{2\pi i z},p=e^{2\pi i\tau}, q=e^{2\pi i \sigma}$, it is then defined by
\begin{equation}\label{eq:def-2q-pochhammersymbol}
(x;p,q)_\infty = \prod_{j,k=0}^\infty (1-x p^j q^k) \,.
\end{equation}
For $\operatorname{Im}\tau>0$ and $\operatorname{Im}\sigma>0$, the elliptic Gamma function is then defined by \cite{Felder_2000} 
\begin{equation}\label{eq:def-ellipticGamma}
\Gamma(z;\tau,\sigma) = \frac{(\frac{pq}{x};p,q)_\infty}{(x;p,q)_\infty} = \exp\left[ \sum_{l=1}^\infty \frac{1}{l} \frac{x^l -p^lq^lx^{-l}}{(1-p^l)(1-q^l)}\right] \,.
\end{equation}
The product representation makes clear the analogy with the $q$-$\theta$ function, while the plethystic form in the second line provides the natural starting point for the Mellin space construction below.

The elliptic Gamma function satisfies an SL$(3,\mathbb{Z})$ modular relation. 
Due to the $\mathbb{Z}^3$ parametrizing the shifts in $z$, there are many identities involving elliptic Gamma function and $q$-$\theta$ function, yielding various equivalent forms of the SL$(3,\mathbb{Z})$ modular formula \cite{Felder_2000,Jejjala:2022lrm}. 
For the purpose of the present analysis, it is convenient to use the following equivalent form:
\begin{equation}\label{eq:elliptic_gamma_modular}
\Gamma(z;\tau,\sigma) = e^{-i\pi Q(z-1;\tau,\sigma)} \frac{\Gamma(\frac{z-1}{\tau}; -\frac{1}{\tau} , \frac{\sigma}{\tau})}{ \Gamma(\frac{z-\tau-1}{\sigma}; -\frac{\tau}{\sigma} , -\frac{1}{\sigma})} \,.
\end{equation}
Here $Q(z;\tau,\sigma)$ is the cubic polynomial introduced in \eqref{eq:def-Sl3Z-ellipticGamma}. 
Although the infinite-product definition above is initially written for $\operatorname{Im}\tau,\,\operatorname{Im}\sigma>0$, the transformed elliptic Gamma functions in \eqref{eq:elliptic_gamma_modular} are understood through their meromorphic continuation.

Our aim in this section is to show that the three elliptic Gamma functions appearing in \eqref{eq:elliptic_gamma_modular} originate from a single trilinear reflection structure in Mellin space. 
In close analogy with the bilinear construction of the $q$-$\theta$ function, the invariant part of the modular transformation will arise from a reflection identity of the Mellin kernel, whereas the cubic polynomial phase will be generated by the residues crossed in the associated multiple variable contour deformation.

\subsection{SL$(3,\mathbb{Z})$ modularity as trilinear reflection} \label{ssec:triilinearreflection}
\subsubsection{Reflection formula} 

We parametrize the elliptic variable as $z=u+v\tau+w\sigma$,
with $u,v,w\in\mathbb{R}$. 
Since the three directions $1,\tau,\sigma$ are overcomplete in the complex plane, this parametrization is not unique. 
Nevertheless, for any chosen decomposition one may write
\begin{equation}
u=u_0+\lfloor u\rfloor,\qquad v=v_0+\lfloor v\rfloor,\qquad w=w_0+\lfloor w\rfloor,
\end{equation}
where $\lfloor \, \rfloor$ denotes the floor function while $u_0,v_0,w_0\in[0,1)$.
Hence, after these integer shifts $z\rightarrow z-\lfloor u \rfloor-\lfloor v\rfloor \tau-\lfloor w\rfloor\sigma$, the elliptic variable can always be represented with all three coefficients lying in $[0,1)$.
These shifts along the $1$, $\tau$, and $\sigma$ directions are controlled by the periodicity and quasi-periodicity relations of the elliptic Gamma function (see for instance \eqref{eq:multiple-Gamma-recursion}).
It is therefore sufficient to carry out the following analysis in the region $0<u,v,w<1$, with the boundary values excluded to avoid the singular loci appearing in the Mellin representation.
In terms of this parametrization, the elliptic variables appearing in the two modular transformed functions take the form
\begin{align}
\frac{z-1}{\tau} &=v+(1-u)\left(-\frac{1}{\tau}\right) +w\frac{\sigma}{\tau}, \\
\frac{z-\tau-1}{\sigma} &=w+(1-v)\left(-\frac{\tau}{\sigma}\right)
+(1-u)\left(-\frac{1}{\sigma}\right).
\end{align}
As a result, the two modular transformed functions in \eqref{eq:elliptic_gamma_modular} induce the transformations
\begin{align}
(u,v,w;\tau,\sigma) &\longrightarrow
\left(v,1-u,w;-\frac{1}{\tau},\frac{\sigma}{\tau}\right), \\
(u,v,w;\tau,\sigma) &\longrightarrow
\left(w,1-v,1-u;-\frac{\tau}{\sigma},-\frac{1}{\sigma}\right).
\end{align}

We can now write the generalized $q$-Pochhammer symbol in Mellin space by applying the Cahen-Mellin representation \eqref{eq:Cahen--Mellin} independently to the two lattice directions:
\begin{eqnarray}
&& \ln (x;p,q)_{\infty} = - \sum_{m,n=0}^\infty \sum_{l=1}^\infty\frac{1}{l} e^{2\pi i l u} e^{2l(n+v)\pi i \tau} e^{2l(m+w)\pi i \sigma} \\
&=&- \left( \frac{1}{2\pi i}
\right)^2 \int_{c_1 -i\infty}^{c_1+i\infty} ds_1 \int_{c_2 -i\infty}^{c_2+i\infty} ds_2  \, \frac{1}{(-2\pi i \tau)^{s_1}} \frac{1}{(-2\pi i \sigma)^{s_2}}  \\
&& \times \Gamma(s_1) \Gamma(s_2) \text{Li}_{s_1+s_2+1} (e^{2\pi i u}) \zeta(s_1,v) \zeta(s_2,w) \,. \nonumber
\end{eqnarray}
Here the contours may initially be chosen with $c_1,c_2>1$, where the interchange between the sums and the Mellin integrals is justified.
The two independent lattice sums associated with $\tau$ and $\sigma$ give rise to the two Hurwitz zeta functions, while the plethystic $l$-sum produces the polylogarithm of order $1+s_1+s_2$.
Consequently, in contrast to the $q$-$\theta$ function, the elliptic Gamma function naturally requires a double Mellin representation.

Using the definition \eqref{eq:def-ellipticGamma} and the two component vectors introduced in \eqref{eq:def-mathbfLH}, the logarithm of the elliptic Gamma function can therefore be written compactly as 
\begin{align}\label{eq:Mellin-ELGamma-final}
	\begin{split}
\ln \Gamma(z;\tau,\sigma) 
&= \left(\frac{1}{2\pi i} \right)^2 \int_{c_1 -i\infty}^{c_1+i\infty} ds_1 \int_{c_2 -i\infty}^{c_2+i\infty} ds_2  \, \frac{1}{(-2\pi i \tau)^{s_1}} \frac{1}{(-2\pi i\sigma)^{s_2}} \Gamma(s_1) \Gamma(s_2) \\
& \times \varepsilon [\mathbf{L}_{s_1+s_2} (u) \circ \mathbf{H}_{s_1}(v) \circ \mathbf{H}_{s_2} (w)] \,.
	\end{split}
\end{align}
for Im$(\tau)>0$ and Im$(\sigma)>0$. 
Since the componentwise product $\circ$ is both associative and commutative, the Frobenius pairing introduced in the previous section extends naturally to this trilinear contraction, with no preferred ordering among the three vectors. 
Thus, the bilinear Mellin kernel of the $q$-$\theta$ function is promoted to a trilinear structure involving one polylogarithmic vector and two Hurwitz zeta vectors.

The advantage of the $(u,v,w)$ parametrization is that the two modular images can be written in exactly the same trilinear form as the original elliptic Gamma function, with only a permutation and reflection of the arguments of the vectors. 
For the first modular transformed function, assuming additionally that
$\operatorname{Im}(\sigma/\tau)>0$, we obtain
\begin{align}\label{eq:ellipticGamma-modularimage1}
	\begin{split}
& \ln \Gamma \left(\frac{z-1}{\tau};-\frac{1}{\tau},\frac{\sigma}{\tau} \right)\\
=& \left(\frac{1}{2\pi i} \right)^2 \int_{c_1 -i\infty}^{c_1+i\infty} ds_1 \int_{c_2 -i\infty}^{c_2+i\infty} ds_2  \,  \Gamma(s_1) \Gamma(s_2) \left( \frac{2\pi \sigma}{ i \tau}\right)^{-s_2} \left( \frac{2\pi i}{ \tau}\right)^{-s_1} \\
&\times \varepsilon  \left[\mathbf{L}_{s_1+s_2} (v)  \circ \mathbf{H}_{s_1}(1-u)  \circ \mathbf{H}_{s_2} (w)
		\right] \,,
	\end{split}
\end{align}
and for the second in \eqref{eq:elliptic_gamma_modular}, it is explicitly
\begin{align}\label{eq:secondGammaimage-2}
	\begin{split}
& \ln \Gamma\left( \frac{z-\tau-1}{\sigma};-\frac{\tau}{\sigma},-\frac{1}{\sigma}\right) \\
=& \left(\frac{1}{2\pi i} \right)^2 
\int_{c_1 -i\infty}^{c_1+i\infty} ds_1 \int_{c_2 -i\infty}^{c_2+i\infty} ds_2  \,  \Gamma(s_1) \Gamma(s_2) \left( \frac{2\pi i \tau}{  \sigma}\right)^{-s_1} \left( \frac{2\pi i}{\sigma}\right)^{-s_2} \\
&\times \varepsilon  [\mathbf{L}_{s_1+s_2}(w)  \circ \mathbf{H}_{s_1} (1-v)  \circ
\mathbf{H}_{s_2} (1-u)] \,.
	\end{split}
\end{align}
We see that the three Mellin representations have the same trilinear structure, while the roles of the parameters $(u,v,w)$ are reorganized among the polylogarithmic and Hurwitz zeta vectors. 
Thus, in contrast to the one dimensional Mellin reflection $s\rightarrow -s$ encountered for the $q$-$\theta$ function, the SL$(3,\mathbb Z)$ transformation acts nontrivially on the two dimensional Mellin space. 
To compare the three Mellin kernels, one must therefore combine a linear transformation of $(s_1,s_2)$ with the reflection relations exchanging $\mathbf{L}$ and $\mathbf{H}$.

In order to transform \eqref{eq:ellipticGamma-modularimage1} and
\eqref{eq:secondGammaimage-2} into a form that can be directly compared with the Mellin representation of the original elliptic Gamma function \eqref{eq:Mellin-ELGamma-final}, we perform linear transformations of the two Mellin variables.
For \eqref{eq:ellipticGamma-modularimage1}, in order to recover the scale
$\tau^{-s_1'}\sigma^{-s_2'}$, we use the change of variables 
\begin{equation}
	-s_1' = s_1+s_2, \qquad s_2'=s_2 \,.
\end{equation}
Accordingly, the original contours
$\text{Re}(s_1)=c_1$ and $\text{Re} (s_2)=c_2$ are mapped to
\begin{equation}
\text{Re}(s_1)=-c_1-c_2, \qquad \text{Re}(s_2)=c_2 \,.
\end{equation}
Since the Mellin integration variables are dummy variables, we drop the primes after the change of variables.
This transforms \eqref{eq:ellipticGamma-modularimage1} into 
\begin{align}\label{eq:ellipticGamma-modularimage1-reflect}
	\begin{split}
& \ln \Gamma \left(\frac{z-1}{\tau};-\frac{1}{\tau},\frac{\sigma}{\tau} \right)\\
=& \left(\frac{1}{2\pi i} \right)^2 \int_{-c_1-c_2 -i\infty}^{-c_1-c_2+i\infty} ds_1 \int_{c_2 -i\infty}^{c_2+i\infty} ds_2  \,  \Gamma(-s_1-s_2) \Gamma(s_2) 
(2\pi)^{s_1}  \sigma^{-s_2} \tau^{-s_1} \\
&\times e^{\frac{\pi i}{2}(2s_2+s_1)} \varepsilon  \left[\mathbf{L}_{-s_1} (v)  \circ \mathbf{H}_{-s_1-s_2}(1-u)  \circ \mathbf{H}_{s_2} (w) \right] \,.
	\end{split}
\end{align}
Similarly, for \eqref{eq:secondGammaimage-2}, we need to perform the transformation 
\begin{equation}
	s_1'=s_1, \qquad s_2' = -s_1-s_2 \,.
\end{equation}
Then after dropping the prime again, the second modular transformed function can be translated to
\begin{align}\label{eq:secondGammaimage-2-reflect}
	\begin{split}
& \ln \Gamma\left( \frac{z-\tau-1}{\sigma};-\frac{\tau}{\sigma},-\frac{1}{\sigma}\right) \\
=& \left(\frac{1}{2\pi i} \right)^2 
\int_{c_1 -i\infty}^{c_1+i\infty} ds_1 \int_{-c_1-c_2 -i\infty}^{-c_1-c_2+i\infty} ds_2  \,  \Gamma(s_1) \Gamma(-s_1-s_2)  (2\pi)^{s_2} \tau^{-s_1} \sigma^{-s_2} \\
&\times e^{\frac{\pi i}{2}s_2} \varepsilon  [\mathbf{L}_{-s_2}(w)  \circ \mathbf{H}_{s_1} (1-v)  \circ \mathbf{H}_{-s_1-s_2} (1-u)] \,.
	\end{split}
\end{align}

Now define 
\begin{align}
	\begin{split}
\Xi_3(u,v,w;s_1,s_2)= \frac{\Gamma(s_1) \Gamma(s_2)}{(2\pi)^{s_1+s_2}} \, \varepsilon (\mathbf{H}_{s_1} (v) \circ \mathbf{H}_{s_2} (w) \circ \mathbf{L}_{s_1+s_2} (u) )  \,.
	\end{split}
\end{align}
We refer to scalar functions constructed from one polylogarithmic vector and two Hurwitz zeta vectors as $(1,2)$-type scalar functions.
The linear Hurwitz-polylogarithm reflection relations
\eqref{eq:reflectionLH-eachlinear} are sufficient to close the three Mellin frames into a single trilinear identity.
To see this explicitly, we can first apply \eqref{eq:reflectionLH-eachlinear} to $\Xi_3(v,1-u,w;-s_1-s_2,s_2)$. 
The result is
\begin{eqnarray}\label{eq:first-trilinear-frame}
&&\Xi_3(v,1-u,w;-s_1-s_2,s_2)
\\ && =
	-\frac{1}{2\sin(\pi (s_1+s_2))}
\frac{\Gamma(s_1)\Gamma(s_2)}{(2\pi)^{s_1+s_2}} \,
\varepsilon \left( \mathbf{M}(s_1)\mathbf{H}_{s_1}(v) \circ \mathbf{M}(1+s_1+s_2)\mathbf{L}_{s_1+s_2}(u) \circ
\mathbf{H}_{s_2}(w) \right). \nonumber
\end{eqnarray}
Similarly, the second transformed elliptic Gamma function gives
\begin{eqnarray}	\label{eq:second-trilinear-frame}
&&\Xi_3(w,1-v,1-u;s_1,-s_1-s_2) \\
&& = \frac{1}{2\sin(\pi (s_1+s_2))}
\frac{\Gamma(s_1)\Gamma(s_2)}{(2\pi)^{s_1+s_2}} \,
\varepsilon \left( \sigma_1 \mathbf{H}_{s_1}(v) \circ \mathbf{M}(1+s_1+s_2) \mathbf{L}_{s_1+s_2}(u)\circ \mathbf{M}(s_2)\mathbf{H}_{s_2}(w) \right). \nonumber
\end{eqnarray}

The remaining step is purely algebraic. 
To simplify above formula, we should apply the form of connection matrix \eqref{eq:form-MMatrix} and note the simultaneous application of $\sigma_1$ on counit map is invariant 
\begin{equation}
	\varepsilon(\sigma_1 \mathbf{a}_1 \circ \sigma_1 \mathbf{a}_2 \circ \sigma_1 \mathbf{a}_3) = 	\varepsilon(\mathbf{a}_1 \circ \mathbf{a}_2 \circ  \mathbf{a}_3)
\end{equation}
together with $\sigma_1^2=\mathbbm{1}$,  one finds
\begin{align}\label{eq:trilinear-M-algebra}
&e^{\frac{\pi i}{2}s_2} \varepsilon\!\left[
\mathbf{M}(s_1)\mathbf{H}_{s_1}(v) \circ \mathbf{M}(1+s_1+s_2)\mathbf{L}_{s_1+s_2}(u)
\circ \mathbf H_{s_2}(w) \right] \nonumber\\
&\quad + e^{-\frac{\pi i}{2}s_1} \varepsilon\!\left[ \sigma_1\mathbf H_{s_1}(v)
\circ \mathbf{M}(1+s_1+s_2)\mathbf L_{s_1+s_2}(u) \circ \mathbf{M}(s_2)\mathbf H_{s_2}(w)
\right] \nonumber\\
&= -2\sin\!\left[\pi(s_1+s_2)\right] \varepsilon\!\left[
\mathbf H_{s_1}(v) \circ \mathbf H_{s_2}(w) \circ \mathbf L_{s_1+s_2}(u) \right].
\end{align}
Then the modular property of elliptic Gamma function in Mellin space reduces to
\begin{align}\label{eq:Xi3-reflection}
	\begin{split}
& \Xi_3(u,v,w; s_1,s_2) - e^{\frac{\pi i}{2}s_2} \Xi_3(v,1-u,w; -s_1-s_2,s_2)  \\
& + e^{-\frac{\pi i}{2}s_1} \Xi_3(w,1-v,1-u;s_1,-s_1-s_2) =0 
	\end{split}
\end{align}
We refer to this as the \emph{trilinear reflection formula} of elliptic Gamma function. 
The factors $e^{\pi i s_2/2}$ and $e^{-\pi i s_1/2}$ are the connection multipliers associated with the two reflected Mellin frames.
They arise from the Hurwitz-polylogarithm reflection matrix $\mathbf{M}(s)$ together with the phases generated by the corresponding Mellin variable transformations.

The identity \eqref{eq:Xi3-reflection} is a pointwise relation in the
two dimensional Mellin space.  
At the level of the inverse Mellin transform, more care must be taken since these three terms are integrated over distinct cycles.  
It is useful to denote
\begin{equation}
{\cal C}(a,b) := \left\{ (s_1,s_2)\in\mathbb C^2\,\big|\, \mathrm{Re}(s_1)=a,\;
	\mathrm{Re}(s_2)=b \right\},
\end{equation}
with both vertical contours oriented upward.  
The elliptic Gamma function and the two modular transformed function in the modular formula \eqref{eq:elliptic_gamma_modular} are integrated respectively over
\begin{equation}\label{eq:three-Mellin-cycles}
\mathcal{C}_0=\mathcal{C}(c_1,c_2),\qquad \mathcal{C}_1=\mathcal{C}(-c_1-c_2,c_2),\qquad
\mathcal{C}_2=\mathcal{C}(c_1,-c_1-c_2).
\end{equation}
To make the relation between the three integrals manifest, we introduce the common Mellin prefactor and define
\begin{align}\label{eq:three-Mellin-integrands}
\mathcal{I}_0(s_1,s_2) =& (-i\tau)^{-s_1}(-i\sigma)^{-s_2}
\Xi_3(u,v,w;s_1,s_2), \nonumber\\
\mathcal{I}_1(s_1,s_2) =& (-i\tau)^{-s_1}(-i\sigma)^{-s_2} e^{\frac{\pi i}{2}s_2}
\Xi_3(v,1-u,w;-s_1-s_2,s_2), \nonumber\\
{\cal I}_2(s_1,s_2) = & (-i\tau)^{-s_1}(-i\sigma)^{-s_2} e^{-\frac{\pi i}{2}s_1}
\Xi_3(w,1-v,1-u;s_1,-s_1-s_2).
\end{align}
The trilinear reflection identity \eqref{eq:Xi3-reflection} then yields the simple integrand relation
\begin{equation}\label{eq:pointwise-three-integrands}
{\cal I}_0(s_1,s_2) - {\cal I}_1(s_1,s_2) + {\cal I}_2(s_1,s_2) =0 .
\end{equation}

To overcome the difficulty of distinct cycles in \eqref{eq:pointwise-three-integrands},  
we should deform the cycles of two modular transformed function separately to ${\cal C}_0$.  
For the first transformed function, $s_2$ is kept on Re$(s_2)=c_2$, while the $s_1$ contour is shifted from Re$(s_1)=-c_1-c_2$ to Re$(s_1)=c_1$.  
This gives
\begin{align}\label{eq:first-cycle-deformation}
\begin{split}
& \left(\frac{1}{2\pi i}\right)^2 \int_{{\cal C}_1}ds_1ds_2\,\mathcal{I}_1 =
\left(\frac{1}{2\pi i}\right)^2 \int_{{\cal C}_0}ds_1ds_2\,\mathcal{I}_1- \mathcal{R}_1 , \\
& \mathcal{R}_1 = \frac{1}{2\pi i} \int_{c_2-i\infty}^{c_2+i\infty}ds_2
\sum_{\rho_1(s_2)} \underset{s_1=\rho_1(s_2)}{\mathrm{Res}}\, {\cal I}_1(s_1,s_2).
\end{split}
\end{align}
Here the sum runs over the polar divisors crossed by the $s_1$ contour during the deformation.
Similarly, for the second modular transformed function we keep Re$(s_1)=c_1$ fixed and move the $s_2$ contour from Re$(s_2)=-c_1-c_2$ to Re$(s_2)=c_2$.  
Hence
\begin{align}\label{eq:second-cycle-deformation}
\begin{split}
& \left(\frac{1}{2\pi i}\right)^2 \int_{{\cal C}_2}ds_1ds_2\,{\cal I}_2
={} \left(\frac{1}{2\pi i}\right)^2 \int_{{\cal C}_0}ds_1ds_2\,{\cal I}_2-
\mathcal{R}_2 , \\
& \mathcal{R}_2 = \frac{1}{2\pi i} \int_{c_1-i\infty}^{c_1+i\infty}ds_1
\sum_{\rho_2(s_1)} \underset{s_2=\rho_2(s_1)}{\mathrm{Res}}\, {\cal I}_2(s_1,s_2).
\end{split}
\end{align}
After these two deformations, all non-residual contributions are integrated over the same cycle $\mathcal{C}_0$ and therefore cancel by the reflection formula \eqref{eq:pointwise-three-integrands}.  
We consequently obtain
\begin{align}\label{eq:modular-relation-residual}
\mathcal{R}_1-\mathcal{R}_2  =&\ln\Gamma(z;\tau,\sigma) -\ln\Gamma\left( \frac{z-1}{\tau}; -\frac{1}{\tau}, \frac{\sigma}{\tau} \right)
+\ln\Gamma\left( \frac{z-\tau-1}{\sigma}; -\frac{\tau}{\sigma}, -\frac{1}{\sigma}
\right) 
\end{align}
Because the crossed singularities in the double Mellin space include polar divisors depending on $s_1+s_2$, the residual terms $\mathcal{R}_1$ and $\mathcal{R}_2$ are essentially one dimensional contour integrals.  
Further contour deformation localizes them onto the intersections of the polar divisors.  
In the following subsection, we show that $\mathcal{R}_1-\mathcal{R}_2$ reduces to a finite sum of multiple residues and reproduces the cubic modular anomaly.

\subsubsection{Residue and anomaly polynomial}
\label{sssec:Gammaresidue}

The contour representation above leaves the modular anomaly in the form $\mathcal{R}_1-\mathcal{R}_2$. 
In the present problem these residual terms can be evaluated without invoking the general machinery of multivariable residues. 
The essential simplification is that, for $u\notin\mathbb{Z}$, the polylogarithmic factors $\operatorname{Li}_{1+s_1+s_2}(e^{\pm 2\pi i u})$ are holomorphic functions of the Mellin variables. 
Consequently, the polar structure of the original Mellin integrand is generated entirely by the two independent factors
$\Gamma(s_1)\zeta(s_1,\cdot)\,,\Gamma(s_2)\zeta(s_2,\cdot)$. 
The corresponding polar divisors are therefore of normal crossing type, and the ordered contour deformation may be evaluated iteratively.
The pole at $s=1$ originates from the Hurwitz zeta function, while the poles at $s=-n$, with $n\geq0$, originate from the Gamma function.
At the latter points, the values of the Hurwitz zeta function reduce to Bernoulli polynomials through \eqref{eq:Hurwitzzeta-to-Bernoulli}.
Since the dependence coupling $s_1$ and $s_2$ is holomorphic, taking the two residues successively is equivalent to evaluating the local two dimensional residue at an intersection of the corresponding polar divisors.

After carrying out one Mellin integration and summing over the remaining lattice variables, the two residual contributions can be reorganized into logarithms and derivatives of the $q$-$\theta$ function.  
Their difference takes the compact form
\begin{align}\label{eq:R12-theta-resummation}
&{\cal R}_1-{\cal R}_2 = \tau\int^{\,v+w\frac{\sigma}{\tau}} dt\,
\bigg[ \ln\theta\left(t;\frac{\sigma}{\tau}\right) -
\ln\theta\left( \frac{\tau}{\sigma}(t-1); -\frac{\tau}{\sigma} \right) \bigg]
\nonumber\\
&+ \sum_{n=0}^{\infty} \frac{B_{n+1}(u)}{(n+1)!\,\tau^n}
\left. \frac{\partial^n}{\partial t^n} \bigg[ \ln\theta\left(t;\frac{\sigma}{\tau}\right) - \ln\theta\left(
\frac{\tau}{\sigma}(t-1); -\frac{\tau}{\sigma} \right) \bigg]
\right|_{t=v+w\frac{\sigma}{\tau}} \nonumber\\
&+R(\tau,\sigma).
\end{align}
Here $R(\tau,\sigma)$ is an integration constant independent of $v$ and $w$. 
Its value is fixed by matching the resummed expression to the original Mellin contour representation.

Using the modular relation of the $q$-$\theta$ function, the combination
appearing in \eqref{eq:R12-theta-resummation} satisfies
\begin{equation}\label{eq:theta-anomaly}
\ln\theta\left(t;\frac{\sigma}{\tau}\right)-
\ln\theta\left( \frac{\tau}{\sigma}(t-1); -\frac{\tau}{\sigma} \right) =
-i\pi B\left(t-1;\frac{\sigma}{\tau}\right),
\end{equation}
where $B(z;\tau)$ is anomaly polynomial related to second diagonal Bernoulli polynomial introduced in
\eqref{eq:Stransform-qtheta}, whose derivatives of order $n\geq3$ vanish. 
Therefore the infinite residue series in \eqref{eq:R12-theta-resummation} truncates after the $B_3(u)$ term, and we obtain
\begin{align}\label{eq:finite-Bernoulli-anomaly}
&	{\cal R}_1-{\cal R}_2 =-i\pi\Bigg[ \tau\int^{\,v+w\frac{\sigma}{\tau}}dt\,
B\left(t-1;\frac{\sigma}{\tau}\right)+ B_1(u) B\left( v+w\frac{\sigma}{\tau}-1;
\frac{\sigma}{\tau} \right) \nonumber\\
&+ \frac{B_2(u)}{2\tau} \left. \frac{\partial}{\partial t}
B\left(t-1;\frac{\sigma}{\tau}\right) \right|_{t=v+w\frac{\sigma}{\tau}}+
\frac{B_3(u)}{6\tau^2} \left. \frac{\partial^2}{\partial t^2}
B\left(t-1;\frac{\sigma}{\tau}\right) \right|_{t=v+w\frac{\sigma}{\tau}}-
\frac{\tau+\sigma}{12} \Bigg].
\end{align}
The last term is fixed by matching the resummed expression to
the original Mellin contour representation.
Using $z=u+v\tau+w\sigma$, the elementary integral and derivatives in
\eqref{eq:finite-Bernoulli-anomaly} combine precisely into the cubic polynomial appearing in the elliptic Gamma modular transformation,
\begin{equation}\label{eq:R12-final}
{\cal R}_1-{\cal R}_2 = -i\pi Q(z-1;\tau,\sigma).
\end{equation}

We should emphasize that the same anomaly polynomial can also be obtained directly from the multivariable residues of the Mellin integrand $\mathcal{I}_0$ \cite{Zhang:2016kfo}.  
In the present problem this equivalence is particularly transparent.  
Since the polylogarithmic factor depends on $(s_1,s_2)$ only through the holomorphic combination $s_1+s_2$, all singular factors are separated between the two Mellin variables, and the polar divisors are therefore generated independently by the $s_1$- and $s_2$-dependent factors involving the Gamma function and the Hurwitz zeta function.  
Consequently, the iterated one dimensional residues used above coincide with the local multiple residues of the double Mellin integrand at the corresponding intersections of the polar divisors.  
The iterative derivation and the direct multivariate residue calculation thus provide two equivalent descriptions of the same contour contribution, both reproducing the cubic anomaly polynomial.

\subsubsection{Comments}
There is some freedom in applying the Cahen-Mellin representation to
higher-rank elliptic Gamma functions.  
In the construction adopted in this paper, we apply the Mellin representation separately to the two exponential factors associated with the $\tau$ and $\sigma$ lattice directions.  
Together with the parametrization $z=u+v\tau+w\sigma$ this leads naturally to the double Mellin kernel
\begin{equation}
\Gamma(s_1)\Gamma(s_2) \text{Li}_{1+s_1+s_2}(e^{2\pi i u})
	\zeta(s_1,v)\zeta(s_2,w),
\end{equation}
together with its reflected counterpart.  
This representation is particularly well adapted to the Hurwitz-polylogarithm reflection relations developed above and makes the trilinear structure in Mellin
space manifest.

An alternative is to apply the Cahen-Mellin representation directly to the combined exponential
$ \exp\left\{ 2\pi i l\left[ \tau(n+v)+\sigma(m+w) \right] \right\}$. 
The two lattice sums are then combined into a Barnes double zeta function
\cite{komori2010barnes,ruijsenaars2000barnes,kurokawa2003multiple}.
Related double Barnes zeta representations of four dimensional supersymmetric indices, including the $\mathcal{N}=1$ chiral multiplet whose index is an elliptic Gamma function, were recently discussed in \cite{Nakayama:2025hzr}.
In this representation one finds
\begin{align}\label{eq:Gamma-Barnes-representation}
\ln \Gamma(z;\tau,\sigma) =
-\int_{c-i\infty}^{c+i\infty} ds\, \frac{\Gamma(s)\zeta(s+1)}{(-2\pi i)^{s+1}}
\left[ \zeta_2(s,z;\tau,\sigma)- \zeta_2(s,\tau+\sigma-z;\tau,\sigma)
\right].
\end{align}
This form is more economical in the sense that it involves only a single Mellin variable. 
However, we have not found a comparably simple reflection identity of the Barnes double zeta functions which organizes the three modular transformed functions in the manner of the trilinear relation derived above.

The distinction between the two Mellin representations is also relevant
for the analytic structure of the contour deformation.  
In the double Mellin representation used throughout this section, the plethystic
$l$-sum produces
$\text{Li}_{1+s_1+s_2}(e^{2\pi i u})$ rather than a factor such as $\zeta(1+s_1+s_2)$.  
For $u\notin\mathbb Z$, the former is holomorphic in the Mellin variables.  
The singularities of the integrand are therefore carried independently by
$\Gamma(s_1)\zeta(s_1,v)$ and $\Gamma(s_2)\zeta(s_2,w)$.
This separation was crucial in the residue analysis of Section~\ref{sssec:Gammaresidue}: it reduces the multiple residue problem to iterated one dimensional residues and makes the emergence of the cubic anomaly polynomial particularly transparent.

Thus, although the two representations describe the same elliptic Gamma function, they organize its Mellin space information in rather different ways.  
The Barnes representation packages the two lattice directions into a single multiple Barnes zeta function, whereas the double Mellin representation keeps the two lattice directions separate and transfers their coupling to the holomorphic polylogarithmic factor.  
For the purpose of uncovering the reflection structure and the origin of the
modular anomaly, the latter organization appears to be considerably more effective.

\subsection{Transposed trilinear reflection}

The Mellin kernel $\Xi_3$ associated with the elliptic Gamma function contains two Hurwitz zeta vectors and one polylogarithmic vector, and is therefore of $(1,2)$ type in our convention.  
The linear reflection relations in \eqref{eq:reflectionLH-eachlinear}, however, exchange the two basic ingredients $\mathbf H$ and $\mathbf L$ in an almost symmetric manner.  
This suggests that the reflection algebra may also admit a transposed $(2,1)$ sector containing one Hurwitz zeta vector and two polylogarithmic vectors.

Guided by this observation, a short algebraic search leads to the
following combination, which we call the \emph{transpose pair}:
\begin{equation}\label{eq:def-transpose-Xi3}
\tilde{\Xi}_3(u,v,w; s_1,s_2) = \frac{(2\pi)^{-s_1-s_2}}{\Gamma(-s_1-s_2)}\, \varepsilon(\mathbf{H}_{1+s_1+s_2} (u) \circ \mathbf{L}_{s_1} (v) \circ \mathbf{L}_{s_2} (w)) \,.
\end{equation}
Applying the same reflection relations \eqref{eq:reflectionLH-eachlinear} to the
three factors, together with the algebra of the connection matrix $\mathbf{M}(s)$, one finds the trilinear identity
\begin{align}\label{eq:relfection-tildeXi3}
	\begin{split}
& \tilde{\Xi}_3 (u,v,w; s_1,s_2) + e^{\frac{\pi i}{2}(1+s_2)} \tilde{\Xi} (1-v,u,w; -(1+s_1+s_2),s_2) \\
& + e^{\frac{\pi i}{2}(1-s_1)} \tilde{\Xi} (1-w,1-v,u; s_1,-(1+s_1+s_2)) =0  \,.
	\end{split}
\end{align}
The structure is the transpose of the elliptic Gamma kernel: the latter contains two Hurwitz zeta vectors and one polylogarithmic vector, whereas \eqref{eq:def-transpose-Xi3} contains one Hurwitz zeta vector and two polylogarithmic vectors.  
At the same time, the three terms are organized by the same two SL$(3,\mathbb Z)$ transformations of the moduli, accompanied by the corresponding transposed action on $(u,v,w)$.

We now ask whether the transposed Mellin kernel admits a realization of functions in  the $\tau,\sigma$ space analogous to the elliptic Gamma function, and whether such a realization obeys an analogous SL$(3,\mathbb Z)$ transformation law.
Formally, one may associate to it the double inverse Mellin transform: 
\begin{equation}\label{eq:transpose-partition}
\ln\widetilde{\mathcal Z} (u,v,w;\tau,\sigma)=
\left(\frac{1}{2\pi i}\right)^2 \int_{\mathcal C_0} ds_1\,ds_2\,
(-i\tau)^{-s_1} (-i\sigma)^{-s_2} \widetilde{\Xi}_3(u,v,w;s_1,s_2)\,.
\end{equation}
See \cite{doi:10.1142/1425} for examples of double Mellin-Barnes integrals. 
However, there is an important difference from the elliptic Gamma case.  
The Mellin representation of the latter contains independent Cahen-Mellin factors $\Gamma(s_1)\Gamma(s_2)$, which convert the two Mellin integrations directly into elementary exponential kernels.
No analogous factorized Gamma function structure is present in \eqref{eq:def-transpose-Xi3}.  
Indeed, using \eqref{eq:reflectionLH-eachlinear}, the factor $1/\Gamma(-s_1-s_2)$ can be combined with the Hurwitz zeta vector so that the transpose pair is equivalently expressed entirely in terms of polylogarithmic vectors.  
The standard Cahen-Mellin inversion therefore does not reduce \eqref{eq:transpose-partition} to a simple double exponential kernel.

This distinction is already visible at the level of the analytic structure.  
The apparent pole at $s_1+s_2=0$ is removable, and the transpose kernel has no finite Mellin poles for generic $u,v,w\notin\mathbb Z$.  
Hence no residue anomaly analogous to the cubic elliptic Gamma phase is suggested by the finite pole structure.
On the other hand, the absence of independent Gamma function damping also means that the convergence of the ordinary vertical contour inverse Mellin transform is no longer automatic.  
Consequently, \eqref{eq:transpose-partition} should at this stage be regarded as a
formal Mellin space reconstruction rather than an established infinite product representation.

The transposed reflection \eqref{eq:relfection-tildeXi3} thus provides a well-defined and nontrivial Mellin space identity, while its realization in moduli variables appears to be more complicated than that of the elliptic Gamma function.  
It may require a generalized inverse Mellin prescription, or equivalently an
infinite order or non-local operation \cite{cardy1991operator} in the original variables.
Determining whether such a realization can be expressed in terms of a recognizable special function, and understanding its associated SL$(3,\mathbb Z)$ transformation law, are interesting questions that we leave for future work.

\section{Discussion and outlook}
\label{sec:conclusion}

In this work, we have developed a Mellin space description of modularity based on the reflection relations between the Hurwitz zeta function and the polylogarithm. 
The $\theta(z;\tau)$ and $\Gamma(z;\tau,\sigma)$ provide two nontrivial realizations of the same mechanism.
After organizing the basic functions into the two-component vectors $\mathbf{L}_s$ and $\mathbf H_s$, their modular transformations are encoded respectively by reflection identities of bilinear and trilinear scalar functions.  
In both cases, the modular relation naturally separates into two ingredients: a pointwise reflection identity in Mellin space and a contour deformation whose residues generate the corresponding modular anomaly.  
For $\theta(z;\tau)$ this produces the familiar quadratic phase, while for $\Gamma(z;\tau,\sigma)$ the two dimensional Mellin reflection and its associated contour deformation reproduce the cubic diagonal Bernoulli polynomial.  
From this viewpoint, the passage from SL$(2,\mathbb Z)$ to SL$(3,\mathbb Z)$ modularity is realized as the promotion of the same linear Hurwitz-polylogarithm reflection formula \eqref{eq:Hurwitz-Polylog-reflection} from a bilinear to a trilinear Mellin structure.

The usefulness of this viewpoint is further illustrated by reflection structures beyond these two standard examples.
Already at the bilinear level, shifting the Mellin parameters of the polylogarithmic and Hurwitz zeta vectors simultaneously leads to a new scalar reflection kernel whose inverse Mellin transform gives rise to a weighted $q$-Pochhammer type function with a nontrivial modular transformation. 
At the trilinear level, exchanging the numbers of Hurwitz zeta and polylogarithmic vectors leads to a transposed reflection kernel containing two polylogarithmic vectors and one Hurwitz zeta vector.  
This kernel closes consistently under the reflection relations in Mellin space, although its realization by inverse Mellin transformation is considerably less direct, since the independent Cahen-Mellin Gamma factors present in the elliptic Gamma kernel are absent. 
These examples suggest that the Mellin space reflection structure is not restricted to the standard $q$-$\theta$ and elliptic Gamma functions, and may provide a useful organizing principle for constructing and studying broader classes of modular
special functions.

We also comment on possible generalizations to multiple elliptic Gamma functions of rank $r$ \cite{Nishizawa_2001,NARUKAWA2004247} in appendix \ref{appendix}.
Their natural Mellin space kernels are of $(1,r+1)$ type, containing one
polylogarithmic vector and $(r+1)$ Hurwitz zeta vectors.
A natural broader class consists of $(n,r+2-n)$ contractions containing
$n$ polylogarithmic vectors and $(r+2-n)$ Hurwitz zeta vectors.
An interesting question is whether there exist nontrivial subclasses for which the reflection relations close systematically, and among them, which admit sufficiently simple inverse Mellin realizations.

Based on the structures uncovered in this work, there are several interesting directions for further exploration.

First, one of our original motivations was to understand whether there
exist special functions beyond $\Gamma(z;\tau,\sigma)$ that admit analogous SL$(3,\mathbb Z)$ transformation laws, and whether such functions can arise naturally as partition functions of quantum field theories.  
In two dimensions, modular invariance can be converted into bootstrap constraints by acting on the modular crossing equation with differential functionals evaluated at the self-dual point $\tau=i$ \cite{Hellerman:2009bu, Friedan:2013cba}.  
For SL$(3,\mathbb Z)$ modularity, it is not yet clear what the appropriate analogue of such a bootstrap problem should be in $(\tau,\sigma)$ variables.  
The reflection identities \eqref{eq:Xi3-reflection} and \eqref{eq:relfection-tildeXi3} provide a possible starting point for formulating this question directly in Mellin space.  
In particular, they suggest that more general modular objects may be constructed by combining the Hurwitz-polylogarithm reflection relations with suitable Mellin space multipliers and multilinear contractions.  
It would be interesting to enlarge this construction further by incorporating more general functional relations, for example those involving Fox $H$-functions and Meijer $G$-functions \cite{doi:10.1142/8711}.  

One may also consider alternative classes of multiple zeta functions beyond the Barnes type, such as $\zeta(s,t)=\sum_{m>n\ge1}^\infty m^{-s}n^{-t}$ (See for example \cite{doi:10.1142/9634}).  
An important question is then which of the resulting Mellin space reflection structures admit well-defined realizations in the modular variables and whether any of them can be interpreted as partition functions of quantum field theories.
A complementary direction is provided by real analytic elliptic extensions of elliptic Gamma type functions and effective actions, which are closely related to elliptic dilogarithms, and non-holomorphic Eisenstein series  \cite{Pa_ol_2018,Cabo-Bizet:2019eaf}.
It would be interesting to investigate whether such elliptic extensions also
admit a natural description in terms of Mellin space reflection identities.

This perspective may also provide a more natural framework for applying
machine learning methods to modular identities.
Machine learning techniques have been used to learn identities of the dilogarithm and trilogarithm \cite{Dersy:2022bym,zagier2007dilogarithm, kirillov1995dilogarithm}, and were subsequently extended to simplify modular expressions involving $\theta(z;\tau)$ and $\Gamma(z;\tau,\sigma)$ \cite{Fan:2026ceb}.
The present analysis suggests a more direct strategy.
In Mellin space, a reflection identity determines the transformed functional structure, while the associated contour deformation produces the modular anomaly.  
Thus, searching for reflection identities of $\operatorname{Li}_s$, with $s$ itself a complex Mellin variable, amounts directly to searching for candidate modular transformation laws.
It may therefore be natural to formulate machine learning searches in
terms of Mellin space reflection relations and their multipliers.
A related, more difficult problem is to reconstruct analytic functions
in the modular variables from a given Mellin kernel.  
Numerical reconstruction methods provide one possible starting point
\cite{Kriesten:2023uoi}, while machine learning or symbolic regression
may help identify the corresponding analytic inverse Mellin realizations.
The transposed kernel $\widetilde{\Xi}_3(u,v,w;s_1,s_2)$ in \eqref{eq:transpose-partition} provides a concrete example.

A further natural direction is to embed the present construction into
the more general framework of the Lerch zeta function
\begin{equation}
L(s,a,c) = \sum_{n=0}^\infty \frac{e^{2\pi i n a}}{(n+c)^s} \,. 
\end{equation}
Both the Hurwitz zeta function and the polylogarithm arise as special limits of this function, i.e. $L(s,a,1) = e^{-2\pi i a} \text{Li}_{s}(e^{2\pi i a})$ and $L(s,1,c) = \zeta(s,c)$.  
Correspondingly, the Hurwitz-polylogarithm reflection relation
\eqref{eq:Hurwitz-Polylog-reflection} may be viewed as a specialization
of the more general Lerch functional equation \cite{apostol1951lerch},
\begin{equation}\label{eq:L-reflection}
L(1-s,a,c) = \frac{\Gamma(s)}{(2\pi)^{s}} \left[ e^{i\pi \left(\frac{s}{2}-2ac\right)} L(s,1-c,a) + e^{i\pi \left(-\frac{s}{2}+2c(1-a)\right)} L(s,c,1-a) \right] \,.
\end{equation}
See also \cite{lagarias2010lerch,lagarias2010lerch-2,lagarias2016lerch}
for further studies of the analytic and transformation properties of the Lerch zeta function.  
It would be interesting to understand whether the higher order reflection formula in Mellin space developed in the current work can be extended away from the special Hurwitz-polylogarithm strata to the full Lerch function, and whether the resulting reflection structures lead to new special functions with higher rank modular transformation laws.

Another natural direction is to apply the Mellin transform approach to partition functions arising in other counting problems of physical interest.  
The usefulness of this method relies crucially on the existence of a suitable product, or equivalently plethystic, representation of the partition function, from which an associated Dirichlet series can be constructed.  
A particularly interesting but more complicated example is the refined counting partition function of $d$-matrix theory \cite{Aharony:2003sx,Collins:2008gc, Bhattacharyya:2008rb, Ramgoolam:2018epz,Lei:2026fep}.
For $d=2$, the partition function admits the two equivalent representations
\begin{equation}
	\mathcal{Z}(x,y) = \prod_{i=1}^\infty \frac{1}{1-x^i -y^i} = \prod_{\substack{i_1,i_2 =0 \\ i_1+i_2\neq 0}}^\infty  \frac{1}{(1-x^{i_1} y^{i_2})^{a(i_1,i_2)}} \,,
\end{equation}
where given $\phi(q)$ as the Euler totient function, the single particle degeneracy $a(i_1,i_2)$ is
\begin{equation}
	a(i_1,i_2) = \frac{1}{i_1+i_2} \sum_{q|(i_1,i_2)} \phi(q)  \frac{(\frac{i_1+i_2}{q})!}{(\frac{i_1}{q})!  (\frac{i_2}{q})!} \,.
\end{equation}
From the Mellin space point of view, the central problem is then to determine the Dirichlet series generated by $a(i_1,i_2)$ and to understand its analytic structure.  
Such a representation may provide a multivariable analogue of the Meinardus' framework \cite{Lucietti:2008cv,andrews1998theory,hwang2001limit} and offer a systematic route to the asymptotic expansion of the corresponding degeneracies, and may also be compared with methods of analytic combinatorics in several variables
\cite{PWM2024}.  
The first subleading corrections were obtained recently by combinatorial methods in \cite{Lei:2026fep}, while a systematic description of higher order perturbative terms, and in particular of possible non-perturbative corrections, is still lacking.  
It would therefore be interesting to investigate whether the Mellin space
methods developed here can provide a complementary and more systematic
approach to this problem.

\section*{Acknowledgements}

We thank Jie Gu, Song He, Vishnu Jejjala, Sam Leuven, Wei Li,  Sanjaye Ramgoolam, Yang Zhang and Xinan Zhou for useful discussions at various stages of this project.
Y.L. thanks School of Quantum, Kavli Institute of Theoretical Science, University of Chinese Academy of Science, the 7th Workshop on Fields and Strings and also the 2nd Joint Summer School on Theoretical High-Energy Physics (2026 Beijing) for hospitality during the completion of this project.
Y.L.\ is supported by the National Natural Science Foundation of China (NSFC) No.12305081 and the international collaboration and communication grant between NSFC and the Royal Society No.W2421035.
ChatGPT was used to assist in improving the clarity of the exposition;
all scientific results and interpretations are those of the authors.

\appendix

\section{Multiple elliptic Gamma functions and higher rank structures}
\label{appendix}

In this appendix, we briefly review the hierarchy of multiple elliptic
Gamma functions $G_r$ and explain how the $q$-$\theta$ and ordinary elliptic
Gamma functions studied in the main text arise as its first two members.  
We will comment on how the Mellin transformation interacts with multiple elliptic Gamma functions.
We follow the notation of \cite{NARUKAWA2004247,Lei:2024oij}.

Let $x=e^{2\pi i z}$ and define the multiple elliptic moduli and fugacities  as:
\begin{align}
	\begin{split}
\underline{q} = (q_0, \cdots, q_r), &\qquad \underline{\tau} = (\tau_0,\cdots,\tau_r) \,, \\
\underline{q}^-(j) = (q_0, \cdots, \check{q}_j, \cdots, q_r), & \qquad \underline{\tau}^-(j) = (\tau_0, \cdots, \check{\tau}_j, \cdots, \tau_r)  \,,\\
\underline{q}[j] = (q_0, \cdots, q_j^{-1}, \cdots, q_r), & \qquad \underline{\tau}[j] = (\tau_0, \cdots, -\tau_j, \cdots, \tau_r) \,, \\
\underline{q}^{-1} = (q_0^{-1},  \cdots, q_r^{-1}), & \qquad -\underline{\tau}= (-\tau_0, \cdots, -\tau_r) \,. 
	\end{split}
\end{align}
and the $\check{\tau}_j$ indicates this component is removed. 
We also define $|\underline{\tau}|=\tau_0+...+\tau_r$.
For $\operatorname{Im}(\tau_j)>0$, the generalized $q$-Pochhammer symbol is defined by
\begin{equation}\label{eq:multiple-qPochhammer}
(x;\underline{q})_{\infty}^{(r)} :=
\prod_{j_0,\ldots,j_r=0}^{\infty} \left( 1-xq_0^{j_0}\cdots q_r^{j_r} \right).
\end{equation}
Its meromorphic continuation to more general values of the moduli can be obtained by the elementary inversion relations of the generalized $q$-Pochhammer symbol \cite{NARUKAWA2004247,Felder_2000}.

The rank $r$ multiple elliptic Gamma function is then defined by
\begin{equation}\label{eq:multiple-elliptic-Gamma}
G_r(z|\underline{\tau}) :=
\left( q_0\cdots q_r x^{-1}; \underline{q}\right)_{\infty}^{(r)}
\left[ (x;\underline{q})_{\infty}^{(r)} \right]^{(-1)^r} .
\end{equation}
Equivalently, its plethystic representation is generated by the single-particle function
\begin{equation} \label{eq:multiple-Gamma-single-letter}
Y_r(z|\underline{\tau}) =
\frac{ (-1)^{r+1}e^{2\pi iz} - e^{2\pi i(|\underline{\tau}|-z)} }{
\prod_{j=0}^{r} \left(1-e^{2\pi i\tau_j}\right) } \,.
\end{equation}
The hierarchy obeys the recursive difference relation and reflection relation as
\begin{align}\label{eq:multiple-Gamma-recursion}
& G_r(z+\tau_j|\underline{\tau}) = G_{r-1}
\left( z|\underline{\tau}^-(j) \right) G_r(z|\underline{\tau}), \\
&G_r(-z|-\underline{\tau}) G_r(z|\underline{\tau}) = 1 .
\end{align}
The first two members of this hierarchy are precisely the special functions considered in the main text,
\begin{equation}\label{eq:G0-G1}
G_0(z|\tau)=\theta(z;\tau), \qquad G_1(z|\tau,\sigma) = \Gamma(z;\tau,\sigma) \,.
\end{equation}
Thus the SL$(2,\mathbb{Z})$ modular transformation of the $q$-$\theta$ function
and the SL$(3,\mathbb Z)$ modular property of the elliptic Gamma function are the first two examples of the modular hierarchy of $G_r$.

There are two useful representations of the modular formula involving multiple elliptic Gamma functions.
To state the general modular relation, it is convenient to introduce homogeneous variables
\begin{equation}\label{eq:homogeneous-multiple-Gamma}
(z|\underline{\tau}) = \left( \frac{\zeta}{\omega_{r+1}} \,\middle|\,
\frac{\omega_0}{\omega_{r+1}}, \ldots, \frac{\omega_{r}}{\omega_{r+1}}
\right), \qquad \underline{\omega}= (\omega_0,\ldots,\omega_{r+1})  \,.
\end{equation}
The multiple Bernoulli polynomials are defined through
\begin{equation}\label{eq:multiple-Bernoulli}
\frac{ t^{r+2}e^{\zeta t} }{ \prod_{j=0}^{r+1} \left(e^{\omega_jt}-1\right)}
= \sum_{n=0}^{\infty} B_{r+2,n} (\zeta|\underline{\omega}) \frac{t^n}{n!}.
\end{equation}
The modular property of the multiple elliptic Gamma function can then be written as \cite{NARUKAWA2004247}
\begin{equation}\label{eq:multiple-Gamma-modularity}
\prod_{k=0}^{r+1} G_r\left( \frac{\zeta}{\omega_k}
\,\middle|\, \frac{\omega_0}{\omega_k}, \ldots, \frac{\widehat{\omega_k}}{\omega_k}, \ldots,
\frac{\omega_{r+1}}{\omega_k} \right) =
\exp\left[ -\frac{2\pi i}{(r+2)!} B_{r+2,r+2} (\zeta|\underline{\omega}) \right] .
\end{equation}
Here the hat means that the corresponding entry is omitted.
The different arguments on the left-hand side are related by SL$(r+2,\mathbb Z)$ transformations acting on the homogeneous coordinates.  
In particular, $r=0$ gives the SL$(2,\mathbb Z)$ transformation of the $q$-$\theta$ function with a quadratic Bernoulli polynomial, whereas $r=1$ reproduces the SL$(3,\mathbb Z)$ elliptic Gamma relation with a cubic Bernoulli polynomial.

The relevance of this hierarchy to the present construction becomes particularly transparent in Mellin space.  
The similar parametrization as the maintext is
\begin{equation}\label{eq:multiple-Gamma-coordinate}
z=u+\underline{v}\cdot\underline{\tau}, \qquad \underline{v} = (v_0,\cdots, v_r) \,,
\end{equation}
and we denote $\underline{s}=(s_0,\cdots,s_r)$ so that $|\underline{s}|=\sum_{i=0}^r s_i$ follows similarly. 
Applying the Cahen--Mellin representation independently to the $(r+1)$ lattice directions produces one Mellin variable $s_j$ for each $\tau_j$.  
The plethystic sum is again collected into a single polylogarithm whose order depends only on the sum of the Mellin variables.  
In the notation of the main text, the resulting kernel has the schematic form with $(r+1)$ Hurwitz zeta functions and one polylogarithm, which motivates  the $(r+2)$-linear scalar
\begin{equation}\label{eq:higher-rank-Xi}
	\begin{split}
\Xi_{r+2} (&u,\underline{v}; \underline{s}) :=
\frac{ \prod_{j=0}^{r}\Gamma(s_j) }{ (2\pi)^{\sum_{j=0}^{r}s_j} }  \, \varepsilon\left[ \mathbf L_{|\underline{s}|}(u)
\circ \mathbf H_{s_0}(v_0) \circ\cdots\circ \mathbf H_{s_r}(v_r) \right]\,.
	\end{split}
\end{equation}
The corresponding inverse Mellin representation takes the schematic form
\begin{equation} \label{eq:multiple-Gamma-Mellin}
\ln G_r(z|\underline{\tau}) = \,(-1)^{r+1} \left(\frac{1}{2\pi i}\right)^{r+1}
\int \prod_{j=0}^{r}ds_j\, (-i\tau_j)^{-s_j}  \Xi_{r+2} (u, \underline{v}; \underline{s})\,,
\end{equation}
with the contours initially chosen in the common domain of absolute convergence.

From this perspective, the two examples worked out explicitly in the main text are simply the first two members of a natural sequence,
\begin{align}\label{eq:multiple-Gamma-Mellin-hierarchy}
G_0: &\qquad \mathbf L_s\circ\mathbf H_s , \\
G_1: &\qquad \mathbf L_{s_1+s_2} \circ\mathbf H_{s_1} \circ\mathbf H_{s_2},\\
G_r: &\qquad \mathbf L_{|\underline{s}|} \circ \mathbf H_{s_0} \circ\cdots\circ
\mathbf H_{s_r}.
\end{align}
The bilinear and trilinear reflection structures encountered in Sections~\ref{sec:qthetananalysis} and \ref{sec:ellipticGamma} may therefore be viewed as the lowest-rank realizations of the Mellin organization associated with
multiple elliptic Gamma functions.

A complementary representation of the multiple elliptic Gamma function, particularly useful for studying its modular properties, is closely related to the multiple sine function $S_r(z|\underline{\tau})$.  
We refer to \cite{kurokawa2003multiple,NARUKAWA2004247} for its definition and
basic properties.  
Theorem 14 of \cite{NARUKAWA2004247} gives the following representation:
\begin{align}\label{eq:product-Gamma-Sine-2}
	\begin{split}
G_r(z|\underline{\tau}) &= \exp\left[ -
\frac{2\pi i}{(r+2)! } B_{r+2,r+2} (z|\underline{\tau},1) \right] \\
&\times \prod_{k=0}^\infty \frac{S_{r+1}(z+k+1|\underline{\tau})^{(-1)^r} S_{r+1}(z-k|\underline{\tau})^{(-1)^r}}{\exp \{ \frac{i\pi}{(r+1)!} [ B_{r+1,r+1} (z+k+1|\underline{\tau}) - B_{r+1,r+1} (z-k|\underline{\tau})	] \}}
	\end{split}
\end{align} 
As shown in \cite{Lei:2024oij}, this representation is particularly useful for extracting the polynomial part of the modular anomaly and the exact high temperature asymptotics of free conformal field theories in even dimensions.  
It is therefore natural to ask whether this multiple sine representation admits an equally transparent interpretation in terms of Mellin space reflection identities.

Multiple elliptic Gamma functions arise naturally in higher dimensional supersymmetric partition functions.  
In particular, $G_1$ is the familiar one-loop building block of four dimensional superconformal indices, while $G_2$ appears in six dimensional supersymmetric
partition functions and superconformal indices \cite{Lockhart:2012vp, Imamura:2012efi, Spiridonov:2012ww}.
More generally, their modular properties admit a geometric interpretation in terms of the factorization of supersymmetric partition functions and the action of higher rank modular groups \cite{Gadde:2020bov}.  
Motivated by the observation that analytic continuation of physical parameters can be naturally realized in supersymmetric quantum field theory \cite{Nekrasov:2023xzm}, the same hierarchy, after appropriate analytic continuation and specialization of its parameters, also appears in free conformal field theory partition functions in higher dimensions \cite{Lei:2024oij}.
Their modular structures furthermore admit extensions associated with more general geometric data, including rational polyhedral cones \cite{Winding:2016wpw, Tizzano:2014roa}.  
More systematic searches for modular identities beyond the standard examples have also been considered recently \cite{Fan:2026ceb}.

The purpose of this appendix is only to exhibit the higher rank hierarchy underlying the two explicit examples studied in the main text.  
Establishing the complete Mellin space reflection algebra of $G_r$ would require identifying all reflected Mellin frames, their connection multipliers, and the corresponding multidimensional contour deformations.  
Moreover, the linear Hurwitz-polylogarithm reflection relations allow sectors considerably more general than the $(1,r+1)$ structure naturally selected by $G_r$.  
In particular, one may consider contractions containing different numbers of
polylogarithmic and Hurwitz zeta vectors.  
Determining which $(n,r+2-n)$-type contractions admit closed reflection relations defines a substantially broader classification problem, which lies beyond the
scope of the present work.

\bibliographystyle{JHEP}
\bibliography{qec}
 
\end{document}